\documentclass[twocolumn,nofootinbib,floatfix]{revtex4}
\usepackage{amsmath,amssymb}
\usepackage{latexsym,graphicx,epsfig}
\usepackage{curves,eepic}
\def\lb{\linebreak[4]}
\newcommand{\be}{\begin{equation}}
\newcommand{\ee}{\end{equation}}
\newcommand{\bes}{\begin{subequations}}
\newcommand{\ees}{\end{subequations}}
\newcommand{\bea}{\begin{eqnarray}}
\newcommand{\eea}{\end{eqnarray}}
\newcommand{\bear}{\begin{equation}\begin{array}}
\newcommand{\eear}[1]{\end{array}\label{#1}\end{equation}}
\def\ba{$$\begin{array}}
\def\ea{\end{array}$$}
\def\bra{$\begin{array}}
 \def\era{\end{array}$}

\newcommand{\fr}[2]{\dfrac{{ #1}}{{ #2}}}

\newcommand{\pa}{\partial}
\newcommand{\la}{\langle}
\newcommand{\ra}{\rangle}
\newcommand{\fn}[1]{\footnote{{\sf #1}}}
\def\vak{{\varkappa}}

\newcommand{\lr}[1]{ \langle #1 \rangle}

\def\cl{\centerline}

\newsavebox{\fmbox}

\newcommand{\fordef}{\stackrel{def}{=}}
\newenvironment{Itemize}{\begin{list}{$\bullet$} %
{\setlength{\topsep}{0.2mm}\setlength{\partopsep}{0.2mm} %
\setlength{\itemsep}{0.2mm}\setlength{\parsep}{0.2mm}}} %
{\end{list}}
\newcounter{enumct}
\newenvironment{Enumerate}{\begin{list}{\arabic{enumct}.} %
{\usecounter{enumct}\setlength{\topsep}{0.2mm} %
\setlength{\partopsep}{0.2mm}\setlength{\itemsep}{0.2mm} %
\setlength{\parsep}{0.2mm}}}{\end{list}}

\newcommand{\bu}{$\bullet$\ }

\usepackage{wrapfig}

\begin{document}
\renewcommand{\tilde}{\widetilde}

\date{}
\title{The evolution of vacuum states and phase transitions in 2HDM during cooling of Universe}
\author{I.~F.~Ginzburg$^{1,2}$, I.~P.~Ivanov$^{1,3}$, K.~A.~Kanishev$^{1,2}$
\\
  {\small $^1$ Sobolev Institute of Mathematics,  Novosibirsk, Russia,}\\
{\small $^2$ Novosibirsk State University,  Novosibirsk, Russia,}\\
  {\small $^3$ IFPA, Universit\'{e} de Li\`{e}ge, Li\`{e}ge, Belgium}
  }

\begin{abstract}
We consider the evolution of the ground state in the Two Higgs Doublet Model
during cooling down of the Universe after the Big Bang.
Different regions in the space of free parameters of this model
correspond to different sequences of thermal phase transitions.
We discuss different paths of thermal evolution and corresponding
evolution of physical properties of the system for different modern values of the parameters.
\end{abstract}

\maketitle

The Standard Model relies on the Higgs mechanism of the
electroweak symmetry breaking (EWSB). Its simplest realization ({\it minimal SM}) is based on a single weak isodoublet of scalar fields,
which couples to the gauge and matter fields and self-interact via the quartic potential. The Two Higgs Doublet Model (2HDM) presents
the simplest extension of the minimal scheme of EWSB, it contains two such doublets \cite{TDLee} and a number of free parameters,
for a review see e.g. \cite{hunter}.
Many new phenomena can take place in this model in different regions of the space of these parameters.

When describing the properties of 2HDM, one first finds the minimum of the Higgs potential and describes its symmetry
properties, then calculates the masses and the interaction of the gauge bosons and of the physical Higgs bosons,
and then studies the Yukawa sector of the model.
The phenomenological richness of the 2HDM is contained in particular in
the fact that the Higgs potential can support minima (or, in general, extrema)
of different nature.  We will refer to them as the {\em phases} of 2HDM.
Upon continuous change of free parameters of the potential, the exact position of the minimum shifts, but the symmetry properties
at the minimum remain unchanged, so that we still stay in a given phase.
There are special points in the space of free parameters, at which this variation of parameters changes the nature of the global minimum,
so that if the system crosses this point, a phase transition occurs. The phase transition is reflected
in a non-analytic behavior of the properties of the vacuum: vacuum expectation values, masses, etc.

At any given set of parameters, several extrema of the Higgs potential can coexist.
Normally, we will be interested in the properties of the vacuum, i.e. the deepest minimum of the potential,
however sometimes it will be useful to trace the evolution of other extrema too.
In particular, we will often calculate the difference in the potential depth between
the vacuum (which we dub ``the vacuum energy'') and a higher lying extremum (``the extremum energy'').

This richness can become even more manifest at non-zero temperature. The effective parameters of
the model evolve with temperature, which can lead to thermal phase transitions,
with important cosmological implications.

This issue has been studied in a number of papers, \cite{Dvali1996,Kanemura,Cottingham1995,Turok1992,Losada1997,Andersen1999}.
All these papers focus on some more or less simple and specific variants of 2HDM and study them within thermal field theory
at different levels of sophistication.
In the present work we complement those papers by carrying out a systematic analysis of {\em all} possibilities offered by 2HDM at finite temperature.
We are interested in three basic questions (for an earlier discussion of these questions see \cite{Gin06}):
\begin{Enumerate}
\item
which possible thermal sequences of phases are allowed in 2HDM?
\item
what physical phenomena can accompany thermal evolution of the system?
\item
which modern, i.e. low-temperature, values of the parameters of the potential correspond to each phase sequence?
\end{Enumerate}
Although we limit ourselves to the first non-trivial temperature corrections, we find novel phenomena in this approximation,
which were overlooked in previous works. Examples include the possibility of having an intermediate charge-breaking vacuum or
new relations between the first-order phase transitions and the symmetries of the potential.

In our analysis we want to stay as general as possible and yet rely on explicit algebraic calculations.
On the one hand, it is known that the most general 2HDM is intractable with straightforward algebra, so other methods must be used to gain some insight
into its properties. On the other hand, it has been recently proved \cite{Ivan1} that
the entire list of possibilities for the thermal evolution of the most general 2HDM is well preserved
in a specific case of a soft $Z_2$-violating 2HDM, which we think is a more plausible candidate for the description of nature than the most general 2HDM.
But in this case, analytical calculations are possible, see e.g. \cite{GK07}, and we will use them to analyze all the sequences of phase transitions
and typical changes of physical properties of the system in detail.

Let us also outline from the start the method that we will use.

First, we limit ourselves to the first approximation of the thermal perturbation theory in the high temperature limit and to the tree approximation for the potential. This discussion is supposed to describe accurately thermal evolution far from the transition points and the phase sequences at sufficiently high temperature.
At low temperatures, at the end of cooling down, the phase evolution of the Universe could have been more intricate than our discussion suggests. Still, we hope that our discussion is of some value even in this region as it provides a minimal list of phenomena possible in thermal 2HDM.

The important advantage of this approximation is that the Gibbs potential has the same generic form as the zero-temperature
Higgs potential, with only mass terms depending on temperature. This will allow us to describe thermal evolution of the Universe
via a point moving in the phase diagram, whose structure we know from the zero-temperature analysis.

Second, we will assume thermodynamical equilibrium in each moment.
This approximation is justified when kinetic phenomena proceed fast enough and are not affected by the relatively slow expansion of the Universe.
Near the second order phase transition points, when the thermodynamical processes become slow, this approximation might break down.

The structure of the paper is the following. In Section 1 we list some features of the 2HDM Lagrangian and present arguments in favor of the softly $Z_2$-violating Lagrangian for description of reality.
Section 2 is devoted to description of the temperature evolution of the parameters of the Lagrangian.
In section 3 we briefly review all possible types of extrema of the potential and describe the phases of 2HDM,
following mainly \cite{GK07}, \cite{Ivan}. In sections 4--6 we consider, in the tree approximation,
main physical properties of all the possible phases in 2HDM, with a special attention to the case of explicitly $CP$-conserving Higgs potential,
and we discuss possible ways the phase transitions could have occurred during cooling down of the Universe.
Section 7 describes the emergent picture in general.

\section{Lagrangian}\label{sectlagr}

The electroweak symmetry breaking (EWSB) via the Higgs mechanism is described with the Lagrangian
\begin{equation}
{ \cal L}={ \cal L}^{SM}_{ gf } +{ \cal L}_H + {\cal L}_Y \,;\quad { \cal L}_H=T-V_H\, .
\label{lagrbas}
\end{equation}
Here, ${\cal L}^{SM}_{gf}$ describes the $SU(2)\times U(1)$ Standard Model interaction of gauge bosons and fermions,
whose form is independent of the realization of the Higgs sector; the Higgs scalar Lagrangian
${\cal L}_H$ contains the kinetic term $T$ and the potential $V_H$, and ${\cal L}_Y$ describes
the Yukawa interactions of fermions with the Higgs scalars.

In the minimal Standard Model (SM) one scalar isodoublet with
hypercharge $Y=1$ is implemented. Here the kinetic term has the standard form
$T=(D_\mu\phi)^\dagger D_\mu\phi$, and  the Higgs potential is
$V=\lambda(\phi^\dagger\phi)^2/2-m^2\phi^\dagger\phi/2$.
The minimum of $V$ gives the vacuum expectation value $v$ via $\la\phi\ra= v/\sqrt{2}=
\sqrt{m^2/2\lambda}$.  In this model there is one physical Higgs boson remaining;
its coupling constants to the gauge bosons are expressed via their masses.
The Yukawa interaction has form
${\cal L}_Y =\sum g_f^{\rm SM}\overline{Q}_L\phi q_R+h.c.$
with $g_f^{\rm SM}=\sqrt{2}m_f/v. $

In the two-Higgs-doublet model the Higgs potential can have a rather complicated form.
To describe it in a concise way,
it is useful to introduce isoscalar bilinear combinations of field operators
\bear{c}
 x_1=\phi_1^\dagger\phi_1,\,\;\; x_2=\phi_2^\dagger\phi_2,\quad \\[2mm]
x_3=\phi_1^\dagger\phi_2\,,\;\;x_{3^*}\equiv
x_3^\dagger=\phi_2^\dagger\phi_1\,.
 \eear{bilinphi}
The Higgs potential of the most general 2HDM is conventionally parameterized as
 \bear{c}
V=V_0-V_2(x_i)+V_4(x_i)\, ;\\[2mm]
V_2(x_i)=M_ix_i\equiv\\\equiv\fr{1}{2}\left[m_{11}^2x_1\!+\!
 m_{22}^2x_2\!+\!\left( m_{12}^2 x_3\!+\!h.c.\right)\right]\,,\\[2mm]
V_4(x_i)=\Lambda_{ij}x_ix_j/2\equiv\\[2mm]
\equiv\fr{\lambda_1x_1^2\! +\!\!\lambda_2x_2^2}{2}\!+\!
\lambda_3x_1x_2\! +\!\lambda_4x_3x_3^\dagger+\\[2mm]
+\!\!\left[\fr{\lambda_5x_3^2}{2} \!+\!
\lambda_6x_1x_3\!+\!\lambda_7 x_2x_3 \!+\!h.c.\right]\! .
 \eear{potential}
At the first glance, this potential contains 14 free parameters:
real $m_{11}^2, m_{22}^2, \lambda_1, \lambda_2, \lambda_3,
\lambda_4$ and complex $m_{12}^2, \lambda_5, \lambda_6,
\lambda_7$. However the same physical content  is described by a Lagrangian that can be obtained from \eqref{potential}
by a general rotation in the $(\phi_1,\,\phi_2)$ space.
This is known as the {\it reparameterization symmetry}.
Therefore, physical observables depend not on all 14 parameters but on a lower number of their combinations
(cf. discussion in \cite{GK05}). An important particular case of
reparameterization transformation is given by the {\it rephasing transformation}:
 \bear{c}
\left\{ \phi_1\to \phi_1,\quad \phi_2\to \phi_2e^{-i\alpha}\right\}\Rightarrow\;\\[2mm]
 \Rightarrow\;m_{12}^2\to m_{12}^2e^{i\alpha},\quad\lambda_5\to\lambda_5e^{2i\alpha}, \;\;etc.
\eear{rephas}

{\bf Kinetic term}. In the general case the renormalizability requires that the kinetic term $T$ is not diagonal but includes a mixed term \cite{Ginrg}
 \bear{c}
T= D_\mu \phi_1^\dagger D_\mu \phi_1+D_\mu \phi_2^\dagger D_\mu \phi_2+\\[2mm]
 +\vak D_\mu \phi_1^\dagger D_\mu\phi_2 +\vak^* D_\mu \phi_2^\dagger D_\mu\phi_1\,.
 \eear{kapterm}
This kinetic mixing can be eliminated by rotation and renormalization of scalar fields. However, in the general case parameters
of this transformation vary with the renormalization scale even at small distances \cite{Ginrg}.
This unattractive feature is absent in the case of exact or softly broken $Z_2$ symmetry, be it explicit or hidden.

{\bf The $\pmb Z_2$ symmetry, exact and broken}. At $m_{12}^2=0$, $\lambda_6=\lambda_7=0$, $\vak=0$ our system has a $Z_2$ symmetry,
i.e. it is invariant under transformations
 \be
\phi_1\to\phi_1\,,\;\;\phi_2\to -\phi_2\quad or\quad \phi_1\to-\phi_1\,,\;\; \phi_2\to \phi_2\,.\label{Z2rot}
 \ee
In order for this property to survive through the perturbation series,
the Yukawa interactions must connect each right-handed fermion to only one scalar field $\phi_1$ or $\phi_2$
(Models I or II for Yukawa sector, see \cite{hunter} for details of the definitions).

By definition, at $\lambda_6=\lambda_7=0$, $\vak=0$, $m_{12}^2\neq 0$ the $Z_2$ symmetry is {\it softly violated}.
In this case the parameters of the Lagrangian are determined unambiguously
up to the phase rotations of $\lambda_5$ and $m_{12}^2$ that keep $\lambda_5^*m_{12}^4$ fixed.
Models I or II for the Yukawa sector keep this property through the perturbations series.

{\it Hard violation of the $Z_2$ symmetry} requires that at least one of quantities $\lambda_6$ or $\lambda_7$
or $\vak$ differs from zero in each basis of the Higgs doublets.
In this case for renormalizability all these coefficients appear via counter-terms,
and the general Yukawa interactions couple each right-handed fermion to both scalar fields, $\phi_1$ and $\phi_2$
(known as Model III for the Yukawa sector).

The general rotation of the Lagrangian of a softly $Z_2$-violating system in the $(\phi_1,\,\phi_2)$ space
induces nonzero $\lambda_6$ and $\lambda_7$, which makes the Lagrangian look as
if hard violation of $Z_2$ were present.
However, specific relations arising among coefficients of the potential prevent hard $Z_2$ violation
(this situation is known as ``hidden softly $Z_2$-violating case''). We do not use such a form below.

{\bf Explicitly $CP$-conserving form of the potential}
is the one that obeys the symmetry $\phi_i\leftrightarrow \phi_i^\dagger$.
It is provided by the condition that all coefficients
in the Higgs Lagrangian \eqref{potential}, \eqref{kapterm} are real.
The general rotation in the $(\phi_1,\,\phi_2)$ space makes this property hidden.

{\bf Natural choice}. It is often the case that a theory containing more than one field with identical quantum numbers arises
in the low-energy limit of a more fundamental theory, in which these fields were components of a single multicomponent object.
Such a theory often contains a higher symmetry, which might be not obvious at low energies.
So, the fields, which had different values of the new quantum number associated with this symmetry, should not mix at
small distances, while large distances symmetry-breaking mixing terms can appear.
If this argument is to be taken seriously for 2HDM, that is, if the quartic interaction and the kinetic terms
are supposed to reflect some new primary symmetry, then we should limit ourselves only with softly $Z_2$-violating models,
because they allow mixing between $\phi_1$ and $\phi_2$ only at large distances.
We think, therefore, that softly $Z_2$-violating 2HDM is a natural, but sufficiently general choice to be considered \cite{Ginrg}.

That is why we consider in detail evolution of the vacuum state for the Lagrangian with softly broken $Z_2$-symmetry
in the explicitly $CP$ conserving case\fn{The potential with hidden soft $Z_2$ symmetry can be transformed to the potential
of form \eqref{Z2potential} by suitable a rotation. Besides, if $CP$-conserving extremum, with no scalar-pseudoscalar mixing, exists,
then there exists a basis in the $(\phi_1,\,\phi_2)$ space, in which the potential has {\it explicitly CP conserving form},
with real $\lambda_i$, $m^2_{ij}$, \cite{GH05}, \cite{GK05}.
This is the reason why we consider the explicitly CP conserving potential \eqref{Z2potential}. }
 \bes\label{Z2potential}\bear{c}
V_2(x_i)=-\fr{1}{2}\left[m_{11}^2x_1\!+\!
m_{22}^2x_2\!+\!m_{12}^2 (x_3\!+\!x_3^\dagger))\right],\\[2mm]
V_4(x_i)=\fr{\lambda_1x_1^2\! +\!\!\lambda_2x_2^2}{2}\!+\!
\lambda_3x_1x_2\! +\!\lambda_4x_3x_3^\dagger + \\[2mm]
+\fr{\lambda_5(x_3^2+x_3^{\dagger 2})}{2},\;\;  \vak=0\,.
 \eear{Z2potential1}
According to \cite{Ivan1}, this is a representative case of the most general 2HDM, in the sense that all phases and phase transitions
that can happen in the most general case can be mapped to phases and transitions in this model.
In particular, the results for the classification of regions in the parameter space and sequences of phase transitions
derived below coincide with those obtained in the general case \cite{Ivan1}.

Although we focus in this work on the scalar sector only, we note that softly broken $Z_2$-symmetric potential
is inherently stable in perturbative theory only if Yukawa sector has a form of Model I or II. Below we assume that is valid.

It will prove useful to denote $\lambda_2/\lambda_1=k^4$ and introduce a special parametrization for the $V_2$ term
 \bear{c}
m_{11}^2= m^2(1-\delta)\,,\quad
m_{22}^2=k^2 m^2(1+\delta)\,,\\[2mm]
m_{12}^2=\mu km^2\,;\quad k\;\fordef\;\sqrt[4]{\lambda_2/\lambda_1\,},
\eear{potparam}
\ees
so that we switch from the triple $m_{11}^2,\, m_{22}^2,\, m_{12}^2$ to the triple $m^2,\, \delta,\, \mu$.
Note also that the type of the minimum realized at given $m_{ij}^2$ depends not on their absolute values,
but on their ratios. Therefore, the parameter $m^2$ does not appear in the classification of phases,
and we will need only to focus on the $(\mu,\, \delta)$ plane below.

To make some equations shorter, we also introduce the following notation for certain combinations of $\lambda_i$:
 \bear{c}
 \lambda_{345}=\lambda_3+\lambda_4+\lambda_5,\qquad
 \tilde{\lambda}_{345}=\lambda_3+\lambda_4-\lambda_5,\\[2mm]
 \Lambda_{345\pm}=\sqrt{\lambda_1\lambda_2}\pm\lambda_{345},\;\;
 \tilde{\Lambda}_{345\pm}=\sqrt{\lambda_1\lambda_2}\pm\tilde{\lambda}_{345}, \\[2mm] \Lambda_{3\pm}= \sqrt{\lambda_1\lambda_2}\pm\lambda_3.
 \eear{lamnot}

{\bf The $\pmb k$ symmetry}.
The quartic potential in \eqref{Z2potential}, in addition to being $Z_2$-symmetric,
is also symmetric under the transformation
\be
\phi_1\leftrightarrow k\,\phi_2\,.\label{ZKsym}
\ee
At $\delta=0$ this symmetry extends to the entire potential. We call this property the {\em $k$-symmetry}.
This $k$-symmetry is not the true symmetry of the Lagrangian since the kinetic term is not invariant under (\ref{ZKsym}), unless $k=1$.
Loop corrections break this symmetry at $k\neq 1$.

{\bf Positivity constraints.} To have a  stable vacuum, the
potential must be positive at large quasi--classical values of
fields $|\phi_k|$ ({\sl {positivity constraints}}) for an arbitrary direction in the $(\phi_1,\phi_2)$ space.
This translates into $V_4>0$ for all non-zero values of the fields,
which places restrictions on possible values of $\lambda_i$.
For the potential \eqref{Z2potential} such restrictions have a simple form
(see e.g. \cite{dema,GIv})
 \bear{c}
 \lambda_1>0, \quad \lambda_2>0,\quad
\Lambda_{3+}>0,\\[2mm] \Lambda_{345+}>0,\quad
\tilde{\Lambda}_{345+}>0.
 \eear{positivsoft}

{\bf A geometric approach} to the most general 2HDM was developed in \cite{Ivan}, which helps
gain some insight into the properties of the model without long cumbersome calculations.
It is applicable to thermal evolution of the general 2HDM too, \cite{Ivan1}, allowing one to see possible regions
for existence of different types of vacuum and sequences of phase transitions.
Since in this work we limit ourselves to the softly $Z_2$-violating potential, we stick to a more conventional way
of direct calculations, based on \cite{GK07},
to track down the properties of the system during thermal evolution.

\section{Temperature dependence}\label{secdep}

\begin{figure}\vspace{-3mm}
\includegraphics{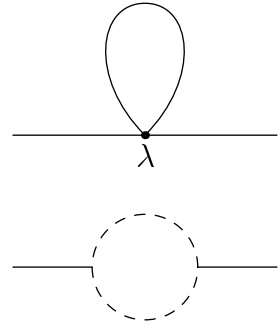}\vspace{-4mm}
\label{figtadpolediag}
\caption{\it 1-st correction to the Gibbs potential}
\end{figure}

At finite temperature, the ground state of a system is given by the minimum of the Gibbs potential, which can be defined as
 \be
V_G= Tr\left(V e^{-\hat{H}/T}\right)/Tr\left(
e^{-\hat{H}/T}\right)\equiv V+\Delta V\,.
 \ee
Corrections  $\Delta V$, to the first nontrivial approximation, are given by diagrams of Fig.~\ref{figtadpolediag}.
Therefore, in this approximation the evolution of the system is described by the potential \eqref{potential}
with fixed quartic term $V_4$ and with the mass term $V_2$ that evolves with temperature.
This contribution is calculated with the Matsubara diagram technique, giving a universal factor,
$T^2/12$, for the loop integral at $T^2\gg m_i^2$ \cite{Tloop}.
The mass terms in the potential \eqref{Z2potential} can be then written as
\bes\label{Tempdep}\bear{c}
 m_{11}^2(T)=  m_{11}^2(0)-2c_1m^2w\,,\\[2mm]
 m_{22}^2(T) =  m_{22}^2(0)-2k^2c_2m^2w\,,\\[2mm]
 m_{12}^2(T)= m_{12}^2(0)\,,\\[2mm]
c_i=c_i^s+c_i^g+c_i^f\,;\qquad w=\fr{T^2}{12m^2}\,.
\eear{Tempdep1}

Here, the scalar loop contributions $c_i^s$ and the gauge boson loop contributions $c_i^g$ (Fig.~\ref{figtadpolediag}, upper diagram) are
 \bear{c}
c_1^s=\fr{3\lambda_1+2\lambda_3+\lambda_4}{2}\,,\quad
c_2^s=\fr{3\lambda_2+2\lambda_3+\lambda_4}{2k^2}\,,\\[2mm] c_1^g=k^2 c_2^g={3 \over 8}(3g^2+g^{\prime 2})\,,
 \eear{scalgaugloop}
with $g$ and $g^\prime$ being the standard electroweak coupling constants.

The fermion loop contributions $c_i^f$ (Fig.~\ref{figtadpolediag}, lower diagram) depend on the form of the Yukawa sector.
For the Model II and Model I, the main contributions to these coefficients can be written in natural notation as
 \bear{c}
c_1^f(II)= {3 \over 2}g_t^2,\quad c_2^f(II)={3 \over 2 k^2}g_b^2\,;\\[2mm] c_1^f(I)={3 \over 2}(g_t^2+g_b^2)\,,\quad c_2^f(I)=0\,.
 \eear{fermloop}
\ees

At given $\lambda_i$ these equations show that {\it the curve of physical states} is the straight ray in the 3-dimensional space
$(m_{11}^2,m_{22}^2,m_{12}^2)$.
This conclusion follows from the universal temperature dependence of the loop integrals
with $\phi_1$ and $\phi_2$ circulating in loops in Fig.~\ref{figtadpolediag}.

The corresponding equations for $c_i^s$ in the most general case are presented in Appendix~\ref{Temphard}.

\subsection{$\lambda_i$ space and $(\mu,\,\delta)$ plane}

The physical state of our system corresponds to some point in the $\lambda_i$ space, whose position is temperature independent,
and a point on the $(\mu,\,\delta)$ plane, which moves as temperature changes.
As temperature increases, this point traces a {\it curve of physical states} on the $(\mu,\,\delta)$ plane,
starting from the {\em ``zero point''}, the point corresponding to the modern, zero-temperature state. Let us describe this curve explicitly.

Simple algebra allows us to express temperature-dependent parameters of the potential   \eqref{potparam}
$m(T)$, $\delta(T)$, $\mu(T)$ via their ``zero-point'' (i.e. modern zero-temperature) values $m$, $\delta$, $\mu$:
 \bear{c}
 m^2(T)=m^2\left[1-(c_2+c_1)w\right]\,,
 \quad \mu(T)=\mu\fr{m^2}{m^2(T)}\,,\\[2mm]
 \delta(T)=\fr{m^2}{m^2(T)}\,
 \left[\delta -(c_2-c_1)w\right]\,,\quad
 .
 \eear{salgexpr}
These relations  are  transformed into equation
\fn{At $k=1$ we have (neglecting fermions) $c_2=c_1\Rightarrow P=0$, i.e. $\delta(T)/\mu(T)=const$.
If additionally $\delta=0$, we have $\delta(T)=0$, i.e. the  $k$-symmetry is a symmetry of the Lagrangian,
and temperature corrections do not break it. Note also that for a purely scalar case
$
P=\fr{k^2-1}{ k^2+1}\cdot\fr{ 3\sqrt{\lambda_1\lambda_2}-2\lambda_3-\lambda_4}{ 3\sqrt{\lambda_1\lambda_2}+2\lambda_3+\lambda_4}\,.
$
},
 \be
 \delta(T)=\fr{\mu(T)}{\mu}\left(\delta-P\right)+P
\,, \quad  P= \fr{c_2-c_1}{c_2+c_1}\,.\label{straight}
 \ee
Eq.~(\ref{straight}) shows that the curve of physical states is a straight ray on the $(\mu,\,\delta)$ plane.
These rays have also the following property: if extended back, beyond the ``zero point'', they all converge
at a single point:
 \be
{\cal P}_0=\left(\mu(T)=0,\,\delta(T)=P\right).\label{crpoint}
 \ee
Of course, this point is physically unattainable, as it would correspond
to $T^2\to -\infty$, however, its identification is useful for visualization of possible thermal trajectories.
Indeed, each ray of physical states can be constructed as starting form a certain ``zero point'' on the $(\delta,\,\mu)$
plane and moving strictly opposite to ${\cal P}_0$.

By virtue of the positivity conditions \eqref{positivsoft} the quantity $c_2+c_1$ is
positive\fn{Indeed, $c_1^s+c_2^s \propto 2\Lambda_{3+} + \Lambda_{345+} + \tilde\Lambda_{345+}$
and $c_i^g>0$, $c_i^f\ge 0$.} so that $m^2(T)$ decreases monotonically.
Therefore, looking back into the past, with temperature increasing, we would see $m^2(T)$ decrease monotonically.
As long as $m^2(T)$ stays positive, the coordinate $\mu(T)$ along this ray grows from today's value $\mu>0$ to infinity,
with $\mu(T)/\mu>0$.
We will depict this evolution on the {\em first sheet} of the $(\mu,\,\delta)$ plane.
At even higher temperatures, at $m^2(T)=0$, we pass onto the {\em second sheet} of
the $(\mu,\,\delta)$ plane, which corresponds to $m^2(T)<0$.
The coordinate $\mu(T)/\mu$ in this ray is negative and grows with temperature rise from $-\infty$ to zero,
while the point on this second sheet of the $(\mu,\,\delta)$ plane moves strictly towards ${\cal P}_0$.

Below, in Figs.~\ref{figewsb}, \ref{fig1delta} and \ref{delmuplane}, we will show these rays of physical states
on the $(\mu,\,\delta)$ plane in different situations. We will see that sometimes
these rays cross regions of different phases, which will have the form of ellipses or line segments.
Directions of these rays, indicated by arrows, always correspond to temperature growth. Small dots, at which these rays start,
indicate various possible ``zero points''. Note that once we have a ray,
we have a freedom to place a ``zero point'' on it, so that any given ray can in fact correspond to several
distinct zero-temperature situations. In all cases we select parameters of the potential
in such a way that $P<0$; rays for positive $P$ can be obtained by appropriate mirror reflections.

\subsection{Sectors in $\lambda_i$ space}

In accordance with the general geometric analysis of \cite{Ivan1},
the entire space of parameters $\lambda_i$ can be divided into sectors
(in this approach each sector corresponds to a relation among eigenvalues of some matrix),
in which different types of vacua are {\bf allowed}, but not necessarily realized:
\begin{equation}
	 \begin{array}{lccc}
		\mbox{Sector I:}  & { \Lambda_{345-}>0}& { \lambda_5 < 0}& { \lambda_4+\lambda_5<0} \\
		\mbox{Sector II:} & { \Lambda_{345-}<0}& \Lambda_{3-} < 0 & \tilde\Lambda_{345-} < 0 \\
		\mbox{Sector III:}&  \lambda_5 > \lambda_4& { \lambda_5 > 0}    & \tilde\Lambda_{345-} > 0\\
		\mbox{Sector IV:} &  \lambda_5 < \lambda_4& \Lambda_{3-} > 0 & { \lambda_4+\lambda_5>0}
  	\end{array}
	\label{sectorstab}
\end{equation}
One can see that sectors do not overlap. Further in the text we
will show how most of these constraints  can be found from detailed analysis of extrema.

In order to know which phase is realized at fixed set of $\lambda_i$ in each sector,
one should look at specific regions on the $(\mu,\,\delta)$-plane.
Thus, description of the phase diagram of 2HDM requires a two-step analysis --- first select $\lambda_i$, then study the $(\mu,\,\delta)$ plane.

\subsection{Evolution of particle masses}

As the Universe was cooling down, the v.e.v.'s of the Higgs fields were changing.
It may have resulted not only in variation of the absolute values of particle masses,
but also in rearrangement of the particle mass spectrum, which can have interesting physical consequences.
To this end, we discuss briefly evolution of masses of the gauge bosons and fermions,
focusing on the Yukawa interaction in the form of Model II, \cite{hunter}.

The mass of $W$-boson is given by $M_W=gv$, while the quark masses are $m_D=g_Dv_1/\sqrt{2}$, $m_U=g_Uv_2/\sqrt{2}$,
where $g$ is electroweak interaction constant, $g_D$ and $g_U$  are Yukawa constants and $U=\{u,\,c,\,t\}$, $D=\{d,\,s,\,b\}$.

During thermal evolution all of $v$, $v_1$, $v_2$ change: $v(T)$, $v_1(T)$, $v_2(T)$.
We express high temperature values of particle masses via their modern values and the ratio of v.e.v.'s to their modern values
\bear{c}
M_W(T)=M_W\fr{v(T)}{v}\,,\\[2mm] m_D(T)=m_D \fr{v_1(T)}{v_1}\,,\quad
m_U(T) =m_U\fr{v_2(T)}{v_2}\,.
\eear{massmotion}
It follows that the ratios between masses of quarks of the same charge in different generations do not change,
e.g. $ m_t(T):m_c(T):m_u(T)= m_t:m_c:m_u$. However, relations between the quark masses of different charge can vary during evolution,
\bear{c}
\fr{m_D(T)}{M_W(T)}=\fr{m_D}{M_W}\,\fr{\cos\beta(T)}{\cos\beta}\,,\,
\fr{m_U(T)}{M_W(T)}=\fr{m_U}{M_W}\,\fr{\sin\beta(T)}{\sin\beta},\\[4mm]
\fr{m_U(T)}{m_D(T)}=\fr{m_U}{m_D}\,\fr{\tan\beta(T)}{\tan\beta}\,.
\eear{massrelmot}
We will show several graphs below that illustrate behavior of the particle spectrum in various situations.

\section{Extrema of potential. Main types}\label{secextrema}

The Higgs potential can have several extrema. The extremum with the lowest value of the energy, the global minimum of potential,
realizes the vacuum state. The other extrema can be either saddle points or maxima or local minima of the potential.

The extrema of the potential define the values $\la\phi_{1,2}\ra$ of
the fields $\phi_{1,2}$ via equations:
\begin{equation}
\partial V/\partial\phi_i
|_{\phi_i=\langle\phi_i\rangle} =0\,,\qquad \partial
V/\partial\phi_i^\dagger |_{\phi_i=\langle\phi_i\rangle} =0\,.\label{vaccondbas}
\end{equation}
These equations have the two main types of solutions:
\begin{Itemize}
\item {\bf electroweak  symmetric (EWs)}
solution $\la\phi_i\ra=0$;
\item  {\bf the electroweak symmetry violating (EWv) solutions} with at least one $\la\phi_i\ra\ne 0$.
 \end{Itemize}
In the next several Sections we will describe these phases and possible thermal phase transitions in the following sequence:
\begin{Itemize}
\item
EWs solution and EWSB phase transitions;
\item
classification of EWv solutions;
\item
evolution of phases in the case when only EWSB phase transition takes place;
\item
evolution of phases in the case of a first-order phase transition between two $CP$-conserving phases;
\item
evolution of phases across $CP$-violating/restoring phase transition;
\item
evolution of phases passing through the charge-breaking phase of 2HDM.
\end{Itemize}

\section{EWs phase. EWSB phase transition}\label{secEWsym}

The EWs point $\la\phi_1\ra=\la\phi_2\ra=0$ is an extremum of the potential. In its vicinity only the mass term $V_2$ can be used for the analysis of the stability.

The EWs point is a minimum of the potential if
\bes\label{ECpoint}
\bear{c}
m_{11}^2<0\,,\;\; m_{22}^2<0\quad\mbox{and}\quad m_{11}^2m_{22}^2\ge |m_{12}^2|^2\,,
\eear{EWmin}
i.~e.  within the circle lying on the second sheet of the $(\mu,\,\delta)$ plane
\be
\mathbb{C}:\;\mu^2+\delta^2=1\,\quad \mbox{with}\quad m^2(T)<0\;\Rightarrow\; \mu(T)/\mu<0\,.\label{circleC}
\ee
The positivity constraints guarantee that this minimum realizes the vacuum\fn{At $\mu(T)/\mu>0$, i.e. on the first sheet of $(\mu,\,\delta)$ plane,
and inside the similar circle \eqref{circleC}, the EWs point is a local maximum of the potential.
For all points outside this circle the EWs extremum is a saddle point.}.
\ees

\begin{figure*}[t]\cl{
\includegraphics[height=4cm]{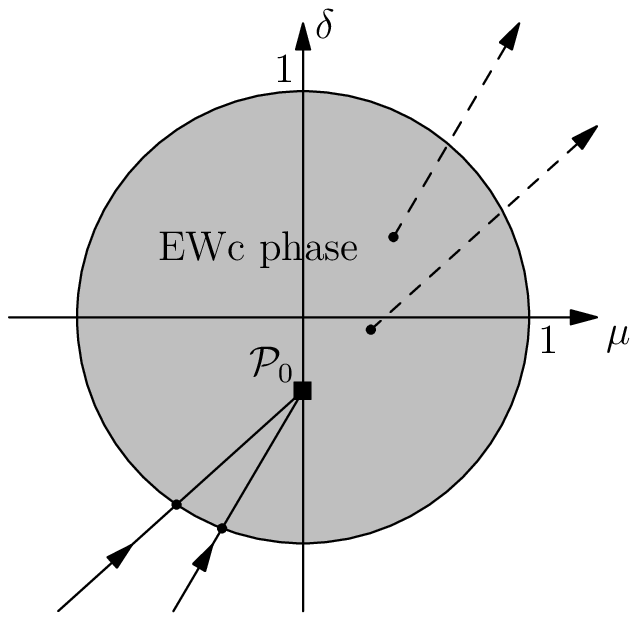}
\includegraphics[height=4cm]{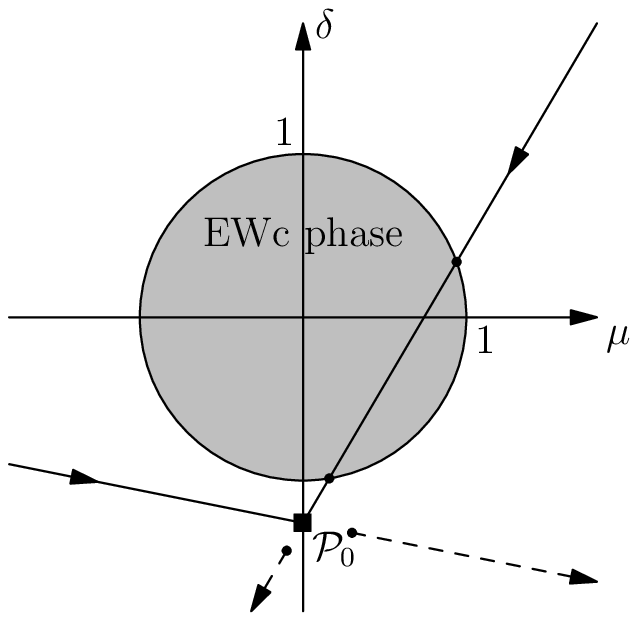}}
\caption{\it The second sheet of the $(\mu,\,\delta)$ plane. Dotted rays are curves of physical states
at the first sheet of this plane, which begin at the modern zero-temperature point
$(\mu_0,\,\delta_0)$ or $(\mu'_0,\,\delta'_0)$. Solid lines are their high-temperature continuations on the second sheet.
The left and right panels correspond to $|P|<1$ and $|P|>1$, respectively.
At points where the solid lines intersect the circle $\mathbb{C}$ the system experiences EWSB phase transitions.}
\label{figewsb}
\end{figure*}

Fig.~\ref{figewsb} shows possible patterns of high-temperature evolution of physical states,
which takes place on the second sheet of the $(\mu,\,\delta)$ plane.
The rays on this second sheet are shown with solid lines; they are the high-temperature continuations of the
rays from the first sheet shown as dotted lines (which are to be described later in Figs.~\ref{fig1delta} and \ref{delmuplane}).
As temperature rises, the point moves towards the special point ${\cal P}_0$, reaching it at infinite temperature.
If it intersects the circle $\mathbb{C}$ $(\mu^2+\delta^2=1)$ at some temperature,
a standard EWSB phase transition takes place.

As can be seen from \eqref{Tempdep}, the evolution of $m_{11}^2$ and $m_{22}^2$ is in general different,
and they change their signs at distinct temperatures. Thus, in contrast with the minimal SM, we can have an intermediate
situation when one of $m^2_{ii}$ is positive, while the other is negative. In this case EWs extremum is a saddle point.

Fig.~\ref{figewsb} shows several scenarios of how the EWSB phase transition could have taken place
at different parameters of the potential.
\begin{Itemize}
\item
If $c_1>0$ and $c_2>0$, then $|P|<1$ and the point ${\cal P}_0$ \eqref{crpoint} lies inside $\mathbb{C}$.
As temperature increases, the point will always cross the circle exactly once on its way towards ${\cal P}_0$.
This happens no matter what the zero-temperature parameters are. This is shown in the left panel of Fig.~\ref{figewsb}.

\item
If $c_1c_2<0$, ${\cal P}_0$ lies outside $\mathbb{C}$.
In this case the system will reside in the EW-violating phase even at extremely high temperature.
With an appropriate choice of modern values of $\mu$, $\delta$,
the ray of physical states can cross the circle $\mathcal{C}$, but {\it it does so twice},
so that the EWs phase becomes an intermediate phase.
This is illustrated in the right panel of Fig.~\ref{figewsb}, ray 2.
Another possibility is that there is no phase transition at high temperature at all,
see ray 1 in the same picture.
\end{Itemize}
{\it In the following sections we discuss only the case with a single EWSB transition, $|P|<1$}.
The analysis of the other possible cases for $m^2(T)<0$ can be conducted in a straightforward way,
but it is decoupled from the discussion of the first sheet of the $(\mu,\,\delta)$ plane.

\section{Electroweak violating (EWv) extrema}

For the EWv solutions we consider also the values of operators $x_i$ at a specific extremum point,
which we denote generically as $N$ (we will omit such subscripts whenever it does not lead to confusion).
 \ba{c}
y_{1,N}\equiv\lr{x_1}_N=\la\phi_1\ra_N^\dagger\,\la\phi_1\ra_N\,,\\
y_{2,N}\equiv\lr{x_2}_N=\la\phi_2\ra_N^\dagger\,\la\phi_2\ra_N\,,\\
y_{3,N}\equiv\lr{x_3}_N=\la\phi_1\ra_N^\dagger\,\la\phi_2\ra_N\,,\\
y_{3,N}^*\equiv\lr{x_3^\dagger}_N=\la\phi_2\ra_N^\dagger\,\la\phi_1\ra_N\,.
 \ea
At each extremum the quantities $y_{1,2}$ are positive, and the Cauchy inequality holds
for an important quantity $Z$:
 \be
y_1>0\,,\quad y_2>0\,,\qquad Z\equiv y_1y_2-y_3^*y_3\ge 0\,. \label{Zcond}
  \ee
Thanks to the properties of homogeneous functions,
we have at each extremum \lb $ V_2(\la \phi_i\ra_N)=-2V_4(\la\phi_i\ra_N)$. Therefore,
the extremum energy is
   \bear{c}
 {\cal E}_N^{ext}=V(\la \phi_i\ra_N)=
 -V_4(\la \phi_i\ra_N)=V_2(\la \phi_i\ra_N)/2\,.
\eear{vacEy}

For each EWv extremum  one can choose the $z$ axis in the
weak isospin space so that
$\la\phi_1\ra={1 \over \sqrt{2}}\begin{pmatrix}0\\v_1\end{pmatrix}$ with real positive $v_1$
(choose ``neutral direction''). In this basis $\la \phi_2\ra$ has
in general an arbitrary form. Then, {\bf after this choice} the most
general electroweak symmetry violating solution of
\eqref{vaccondbas} can be written in a form with real $v_1$ and
complex $v_2$:
 \bear{c}
\langle\phi_1\rangle =\fr{1}{\sqrt{2}}\left(\begin{array}{c} 0\\
v_1\end{array}\right),\quad \langle\phi_2\rangle
=\fr{1}{\sqrt{2}}\left(\begin{array}{c}u \\ v_2
\end{array}\right)\;\;\\[4mm] \mbox{with}\;\; v_1=|v_1|=v\cos\beta,\;v_2=|v_2| e^{i\xi}=v\sin\beta e^{i\xi}\,.
 \eear{genvac}
Without loss of generality one can consider only real positive $u$.

For the most general potential \eqref{potential} one can  classify the EWv extrema
according to their values of $Z$ \cite{GK07}:
\begin{Itemize}
\item If $u\neq 0$, then $Z>0$ --- {\bf charge-breaking extremum}.
\item If $u=0$, then $Z=0$ --- {\bf neutral extremum} with $v^2=2(y_1+y_2)$.
\end{Itemize}

A Higgs potential with arbitrary complex $\lambda_{5,6,7}$, $m_{12}^2$ (i.e. an explicitly $CP$ violating potential)
generally possesses extrema, at which neutral scalars have no definite $CP$ parity (in these states generally $\xi=\arg v_2\neq 0$).
Such potentials in general do not offer any helpful classification of the extrema.

\begin{figure*}[t]
\begin{center}
	 \includegraphics[width=0.45\textwidth]{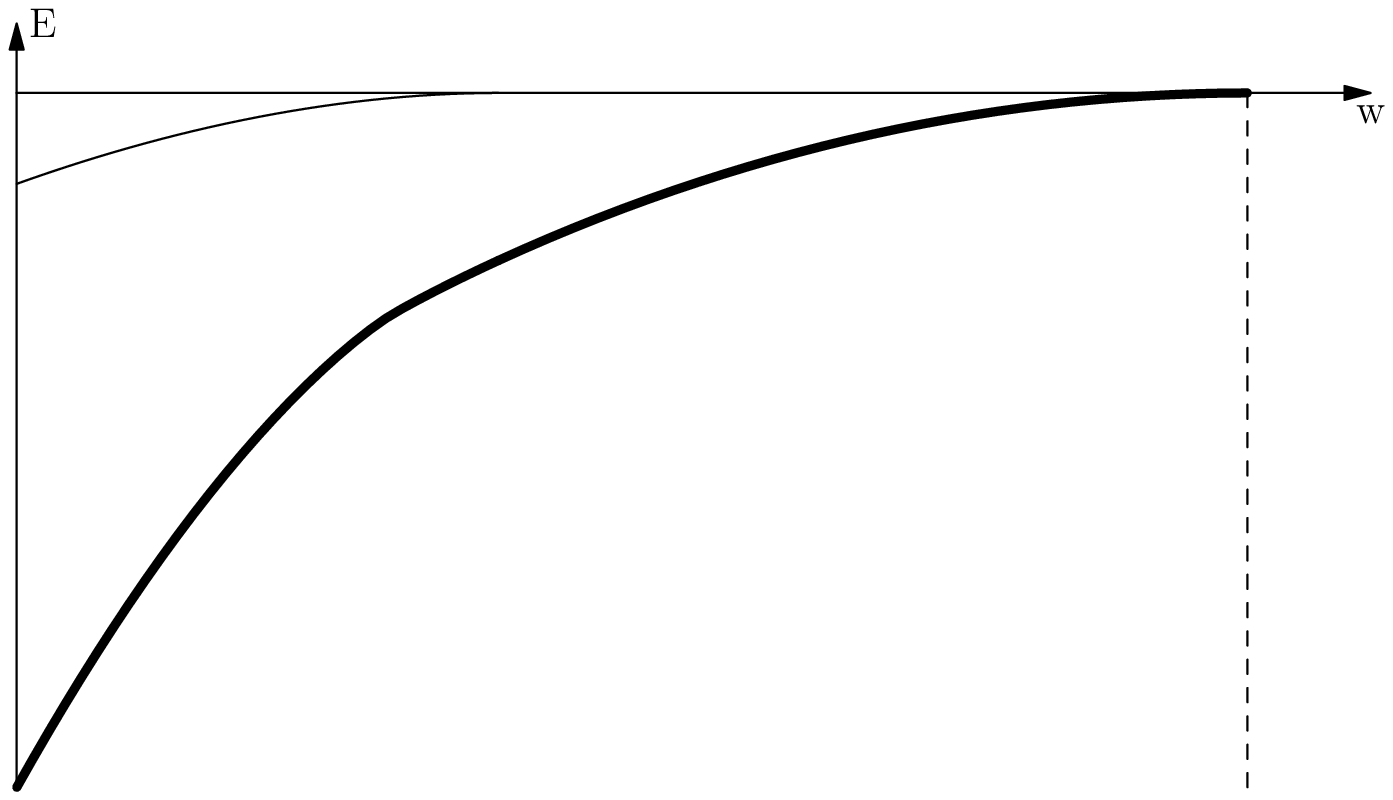}
	\hspace{0.06\textwidth}
	 \includegraphics[width=0.45\textwidth]{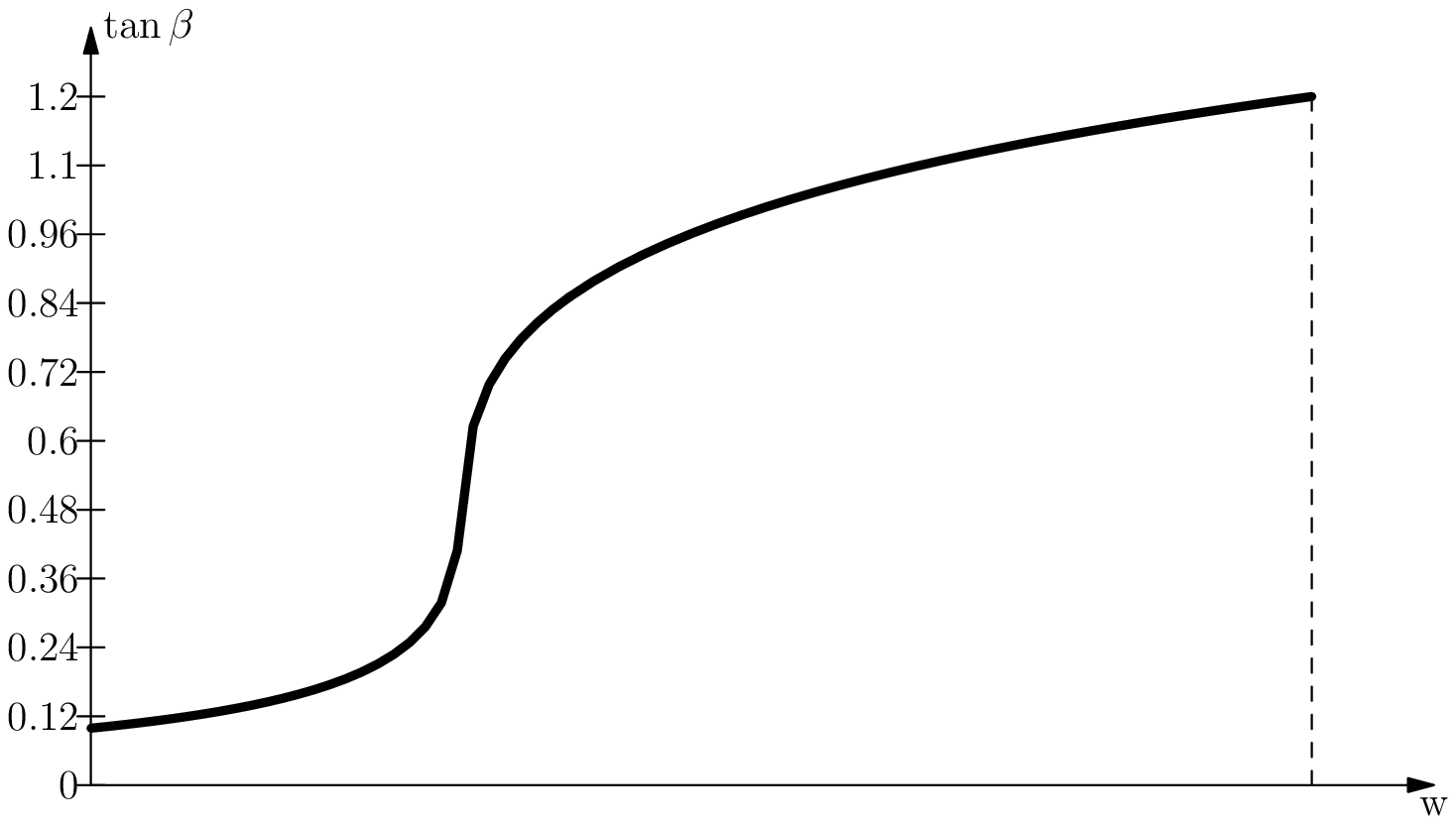}
\end{center}
\includegraphics[width=0.3\textwidth]{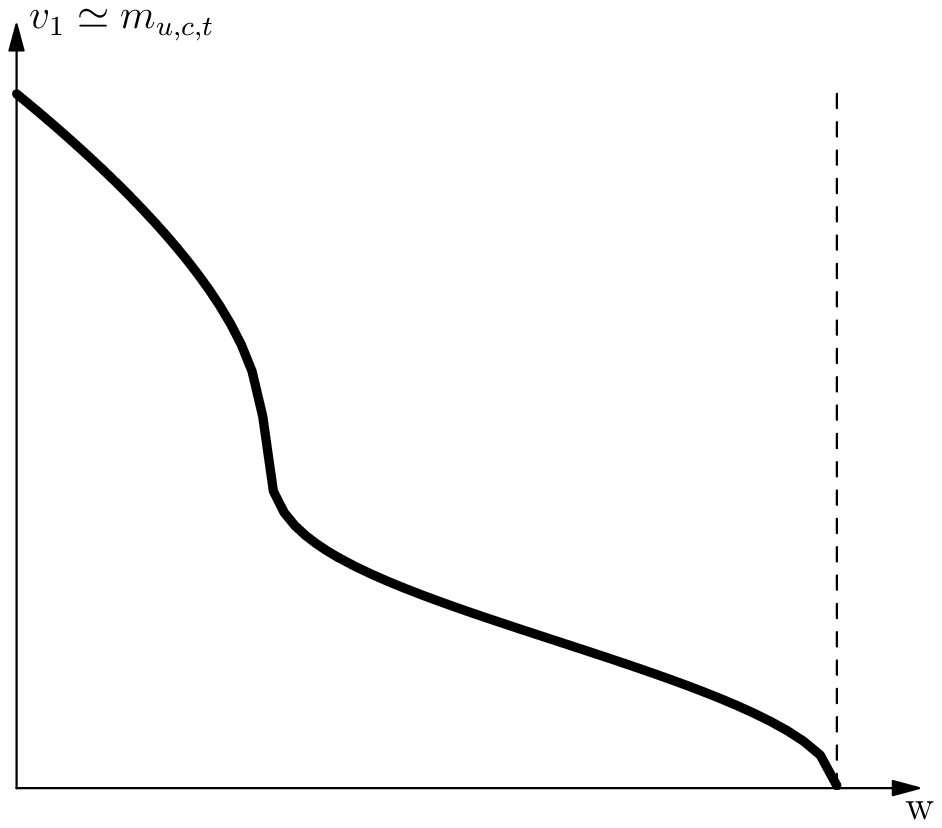}
\includegraphics[width=0.3\textwidth]{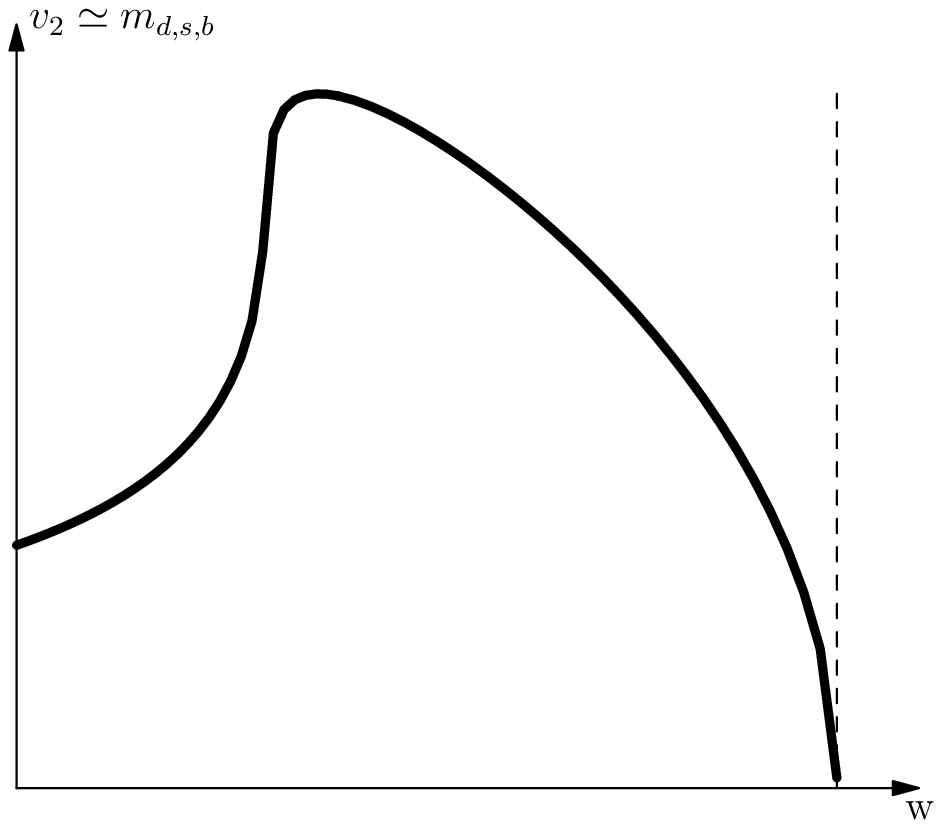}
\includegraphics[width=0.3\textwidth]{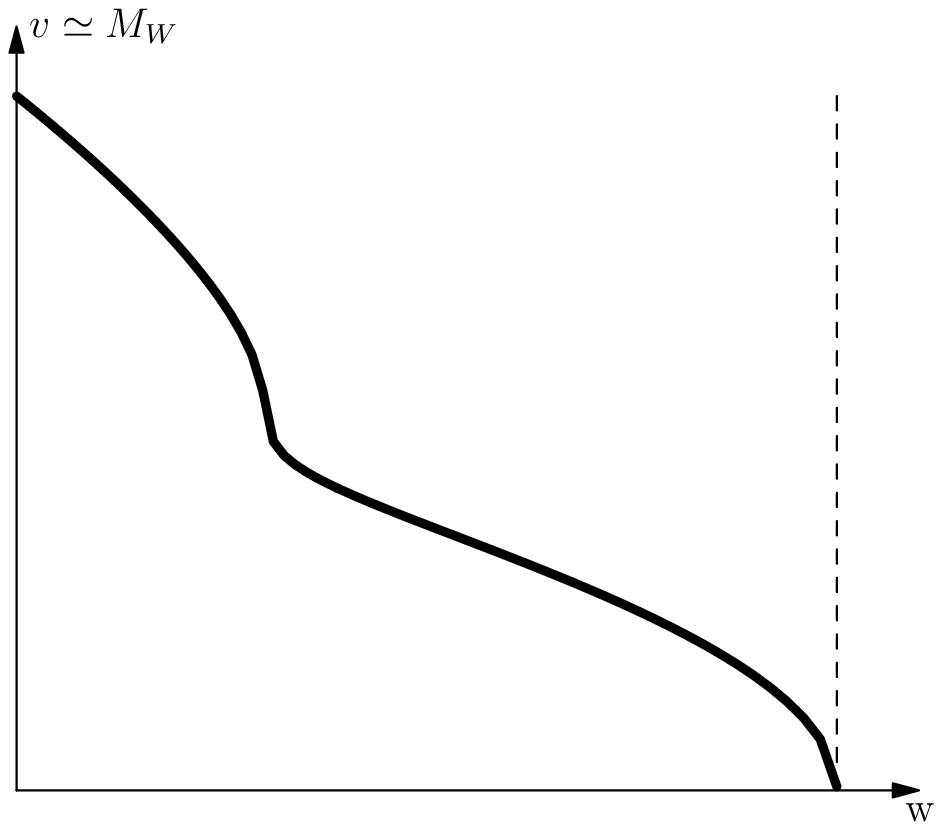}
\caption{\it Parameters of the vacuum state for the case with a single phase transition  EWs$\to$CPc.
Shown are the vacuum and extremum energies (upper left), $\tan\beta$ (upper right),
as well as $v_1$, $v_2$ and $v$ in the lower row.}
\label{1CPcfigA}
\end{figure*}

In the case of explicitly CP conserving potentials, which we focus on in this work,
one can further classify neutral extrema as
 \begin{Itemize}
\item {\it CP conserving --- CPc},
\item  {\it spontaneously CP violating --- sCPv}
 \end{Itemize}
Since there can be up to two $CP$-conserving local minima with different properties,
we will refer to various phases as {\it CPc1, CPc2, sCPv, charge-breaking}.

Two  properties of 2HDM, proven in ref.~\cite{Ivan}, are of much importance for future discussion:
\bes\label{2min}
\bea
	\boxed{
		\begin{array}{c} 
			\mbox{No more than two distinct local } \\ 
			\mbox{minima of potential can coexist.}
		\end{array}}&\quad\label{2coexist}\\
	\boxed{
		\begin{array}{c}
			\mbox{The minima that preserve some symmetry} \\
			\mbox{and that violate it cannot coexist.}
		\end{array}}&\;\label{2differmin}
\eea\ees
Property \eqref{2coexist} means in particular that if
two minima are degenerate, they realize the vacuum state, and the other extrema are not minima.

It follows from the property \eqref{2differmin} that transitions between the states of different symmetry is realized
via merging of these extrema. Therefore all parameters change in this transition continuously,
that is the corresponding phase transition is of the second order
(certainly, taking into account fluctuations might modify this conclusion).

\section{The EWv vacua and phase transitions}\label{secneutrext}

The equations for v.e.v.'s and extremum energies were obtained in \cite{GK07} for each type of extremum. They allow to obtain equations for the sectors
in the parameter space   \eqref{sectorstab}, where certain types of vacuum are realized. Note that transitions among various EWv phases can take place only at $m^2(T)>0$,
i.e. on the first sheet of the $(\mu,\,\delta)$ plane, which are shown in Figures below.

{\bf CPc extrema, general}. The CPc extrema are realized in the entire space of parameters of the potential.
When searching for such an extremum, one can transform two cubic equations representing conditions
\eqref{vaccondbas} into relations for quantities $v^2=2(y_1+y_2)$ and  $\tau=kt\equiv k\tan\beta=k\sqrt{y_2/y_1}$.
For potential \eqref{Z2potential} these equations can be written as
 \bear{c}
v^2=m^2(k^2+\tau^2)\fr{1-\delta+ \mu\tau}
       {\lambda_{345}\tau^2+\sqrt{\lambda_1\lambda_2}}\,,\\[4mm]
\sqrt{\lambda_1\lambda_2}\mu\tau^4+(\Lambda_{345-}-\delta\Lambda_{345+})\tau^3-\\[2mm]
-(\Lambda_{345-}+\delta\Lambda_{345+})\tau-\sqrt{\lambda_1\lambda_2}\mu=0\,.
 \eear{Z2tanbeta}
An algebraic equation of fourth degree
can have 0,~2 or 4 real solutions (including accidental degeneracy).
By construction, one should consider only real solutions of equation \eqref{Z2tanbeta} satisfying $v^2>0$.
Omitting solutions with possible negative values of $v^2$,
one can state carefully that {\it there could be up to 4 CPc extrema}.

The extremum energy is defined via solutions of these equations as
 \be
 {\cal E}_{CPc}=-\fr{m^4k^2}{8}\cdot\fr{(1\!-\!\delta\!+\!\mu\tau)[1-\delta+ 2\mu\tau+\tau^2(1+\delta)]}{\lambda_{345}\tau^2+\sqrt{\lambda_1\lambda_2}}\,.
       \label{CPcenergy}
\ee

Let us now specify different sectors in the $\lambda_i$ space, and discuss the characteristic properties of
thermal evolution of the system in each sector.

\subsection{Sector I. No phase transition except EWSB}\label{figonlyEWSB}

\begin{figure*}[t]
\begin{center}
	 \includegraphics[width=0.45\textwidth]{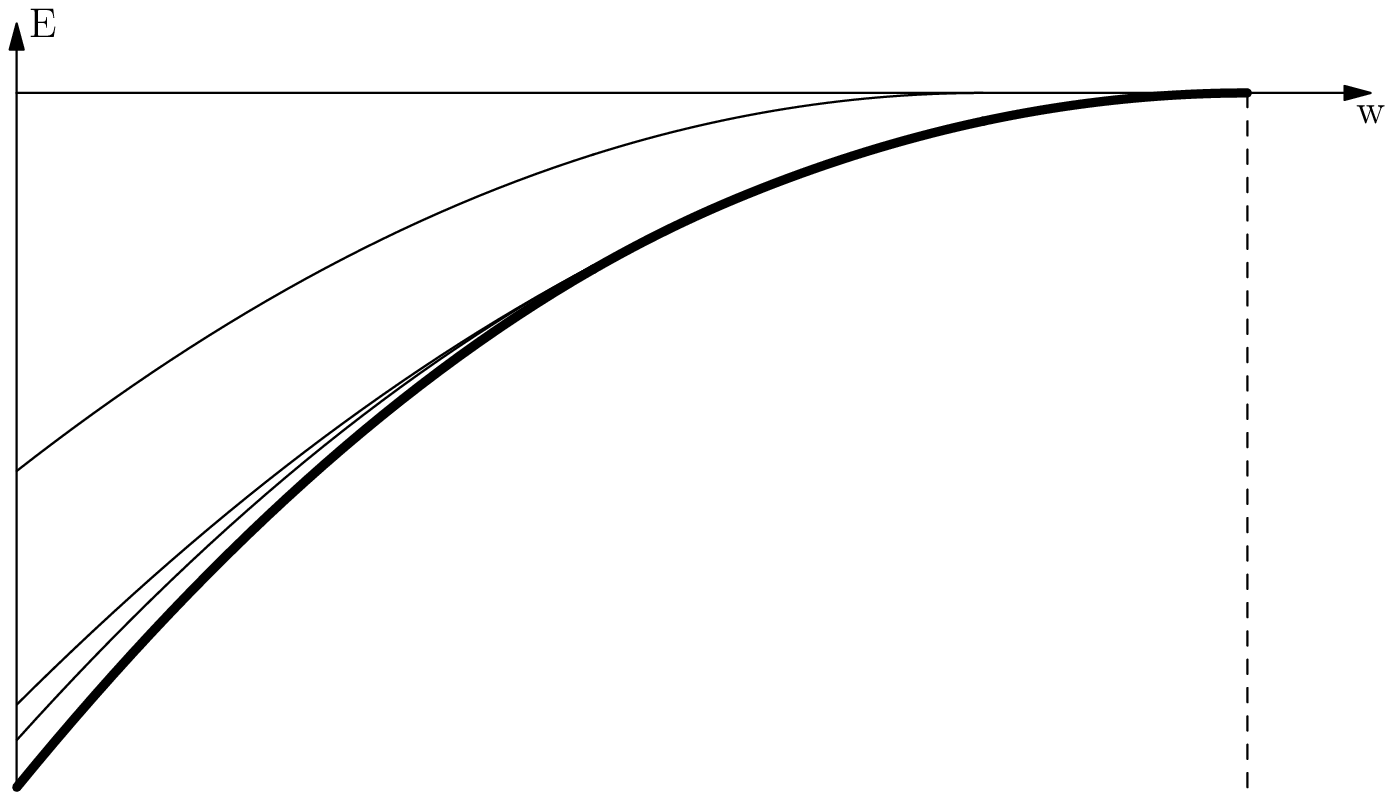}
	\hspace{0.06\textwidth}
	 \includegraphics[width=0.45\textwidth]{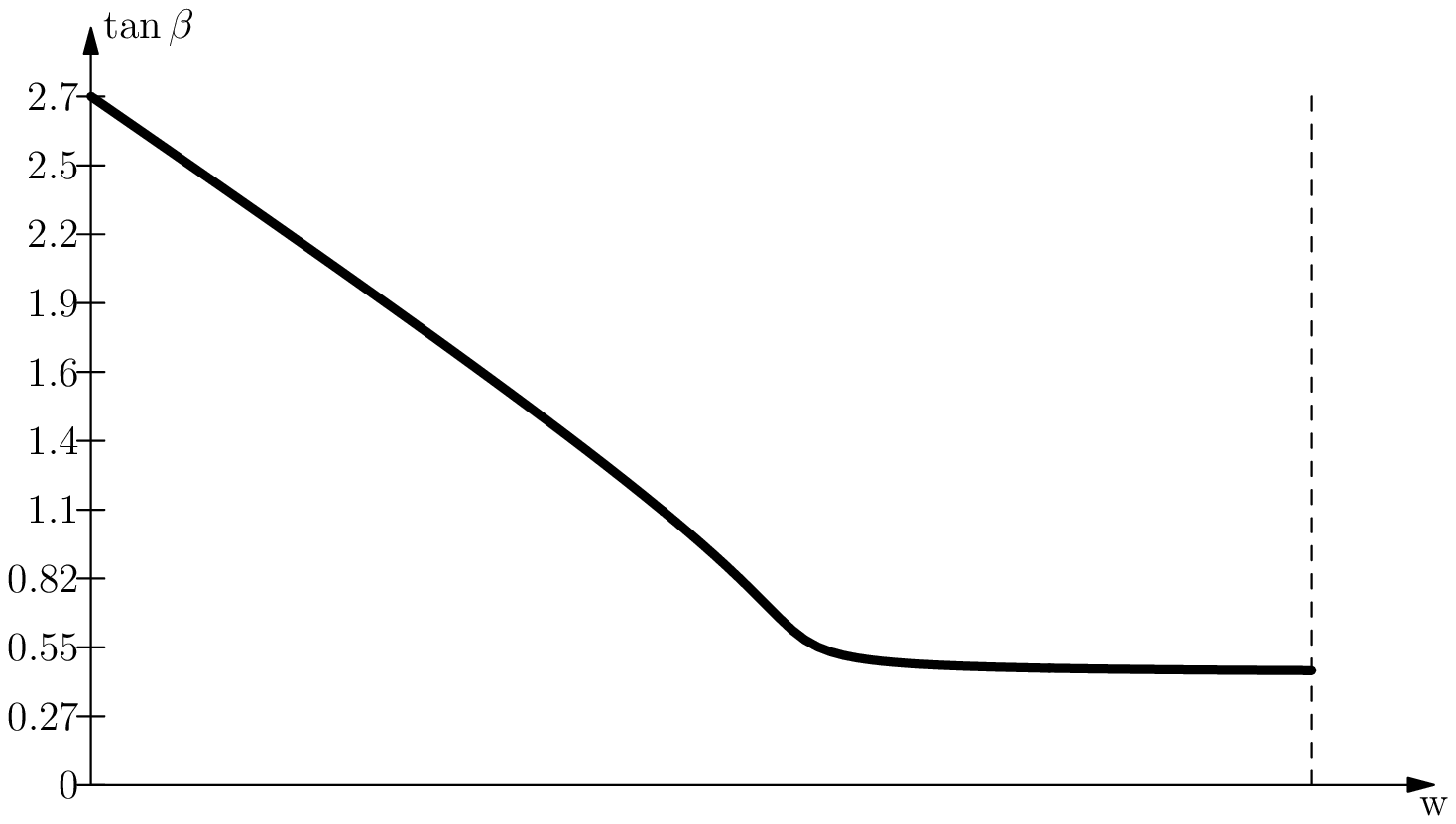}
\end{center}
\includegraphics[width=0.3\textwidth]{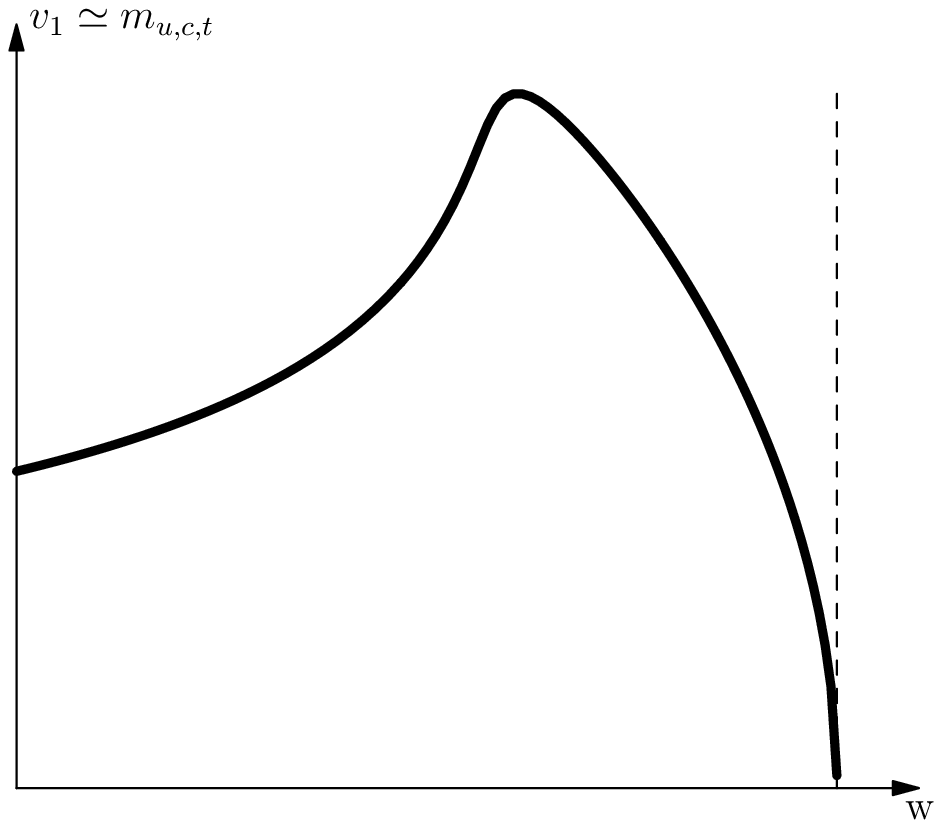}
\hspace{2mm}
\includegraphics[width=0.3\textwidth]{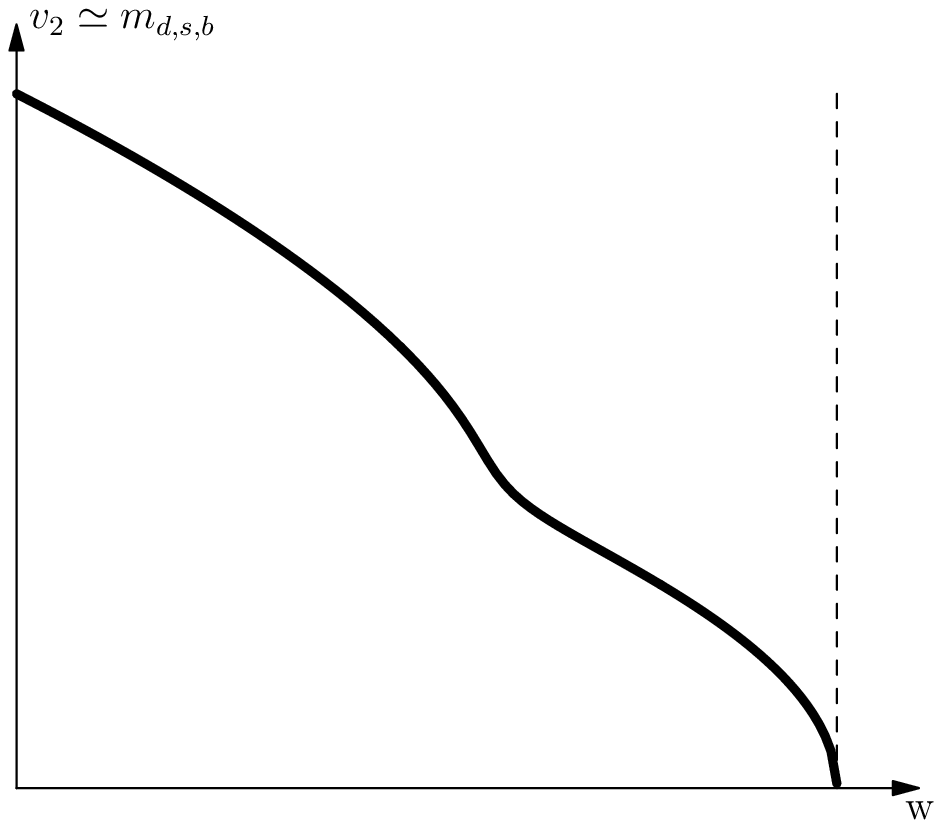}
\hspace{2mm}
\includegraphics[width=0.3\textwidth]{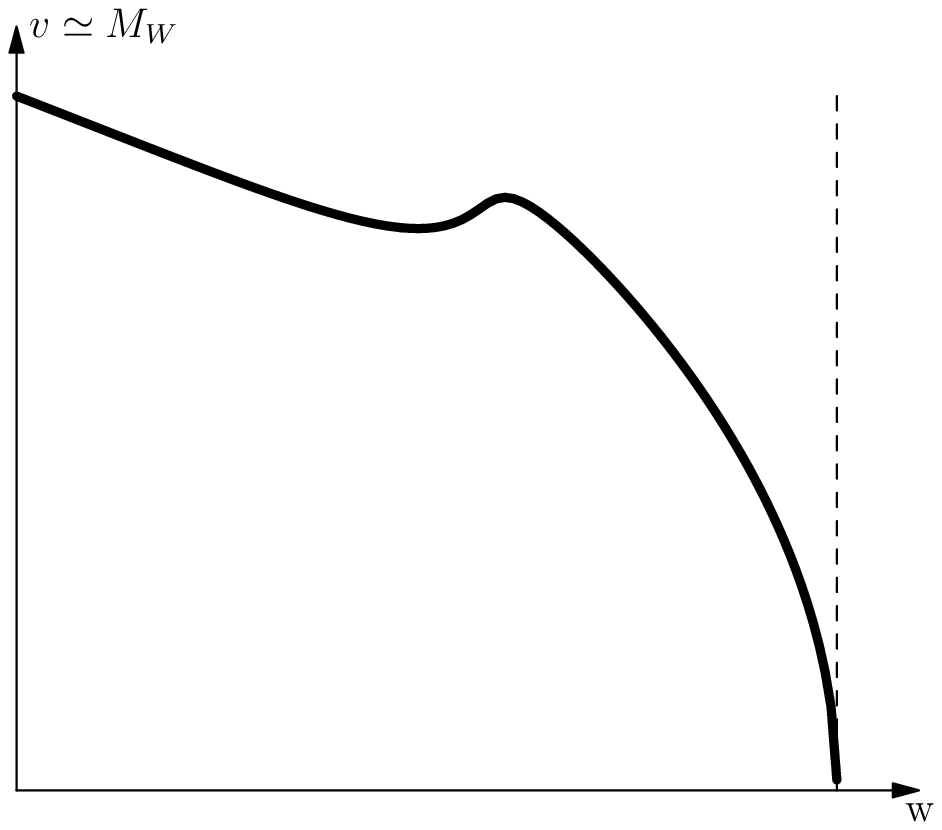}
\caption{\it The same as in Fig.~\ref{1CPcfigA} but for another set of $\lambda_i$.}
\label{1CPcfigB}
\end{figure*}

We start with the case, when there is no other phase transitions apart from the EWSB one.
Such a situation is possible for any $\lambda_i$, within an appropriate region on the $(\mu,\,\delta)$ plane.
Besides, in {\bf sector I} in the $\lambda_i$ space defined by
\be
\lambda_5<0\,,\quad \lambda_4+\lambda_5<0\,,\quad \Lambda_{345-}>0\,,\label{emptysec}
\ee
this situation takes place for any ``zero point'' on the $(\mu,\,\delta)$ plane. These inequalities will be proved below, when we will show that in sectors II, III, and IV several phase transitions can happen, and then (\ref{emptysec}) will appear as a complement to those sectors.

The only phase transition possible in this sector is EWSB transition between the EWs and CPc extrema. Figures \ref{1CPcfigA},~\ref{1CPcfigB} represent, for two different sets of $\lambda_i$, the thermal evolution of phases and their properties in these cases.

Each set of graphs contains evolution of several physical quantities as functions of temperature-dependent parameter $w \propto T^2$. The upper left plot shows the evolution of the vacuum energy (thick curve) as well as the energies of other extrema (thin curves). The dashed line indicates the position of the EWSB transition. The upper right plot gives the evolution of $\tan\beta$. The lower row contains dependencies of $v_1$ and $v_2$, which within Model II for Yukawa sector, give also the temperature dependence of the up-type quark and down-type quark masses. The last plot shows evolution of $v$, which is proportional to the mass of the gauge bosons.

It is remarkable that even without phase transitions, thermal evolution experiences stages of very fast change of the quantities shown. Another remarkable observation is that sometimes masses of the quarks and gauge bosons can depend on temperature in a non-monotonic way. A possibility of rearrangement of the quark mass spectrum is clearly seen from the plots of $v_1$ and $v_2$.

\subsection{Sector II.  First order phase transition between two CPc vacua }\label{sec1ordtr}

A first order phase transition between two $CP$-conserving phases can appear only in the {\bf sector II} in the $\lambda_i$ space,
which is bounded by the following inequalities:
\be
\Lambda_{345-}<0\,,\quad \Lambda_{3-}<0\,,\quad \tilde{\Lambda}_{345-}<0\,.\label{ksymlamlim}
\ee
We explain the origin of these inequalities in this section below.

\begin{figure*}[hbt]
\cl{\includegraphics[width=0.28\textwidth]{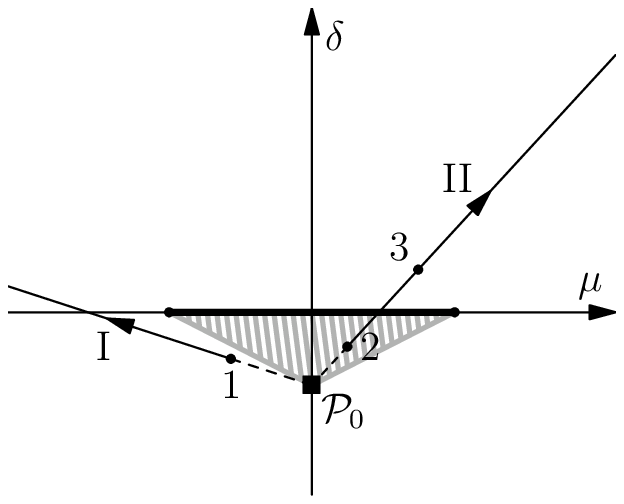}}
\caption{\it The $(\mu,\,\delta)$ plane for the sector II}
  \label{fig1delta}
\end{figure*}

A phase transition between two phases of identical symmetry implies that at some temperature these phases are degenerate.
It was proved in \cite{Ivan} that such a degeneracy always appears in the tree level 2HDM potential
as a result of spontaneous breaking of an extra symmetry of the potential (and not necessarily of the
whole Lagrangian)\fn{The explicit form of the $k$-symmetry changes under reparametrization into more complex linear transformation.}.
In our case this is the $k$-symmetry, which we introduced before.
One can prove that there are no other equally generic symmetries possible.

\bu {\bf Degenerate case with $k$-symmetry  \eqref{ZKsym}} is realized at $\delta=0$ in \eqref{Z2potential}.
In this case the equation  for $\tau$ \eqref{Z2tanbeta} breaks in two independent equations
 $$
 (\tau^2-1)\left(\sqrt{\lambda_1\lambda_2}\mu(\tau^2+1)+\Lambda_{345-}\tau\right)=0
 $$
which can be easily solved.

$\pmb A$. The first of these equations yields {\bf two non-degenerate solutions $\pmb A_\pm$}:
  \bear{c}
(A):\;\tau^2=1\;\Rightarrow\; \tau=\tau_A=\pm 1\,, \\[2mm]
 {\cal E}_{CPcA}=-\fr{m^4k^2}{4}\cdot\fr{(1\pm \mu)^2}{\Lambda_{345+}}\,.
  \eear{Apm}

$\pmb B$.   The second equation yields {\bf two degenerate solutions $\pmb B_\pm$}:
\bear{c}
(B):\; \tau^2+\fr{\Lambda_{345-}}{\mu\sqrt{\lambda_1\lambda_2}}\tau+1=0\;\Rightarrow\\[2mm]
\tau= \tau_{B_\pm}=\fr{-\Lambda_{345-}}{2\mu \sqrt{\lambda_1\lambda_2}}(1\pm R)\,,\quad R=\sqrt{1-\fr{4\mu^2\lambda_1\lambda_2}{\Lambda_{345-}^2}}\,;
 \eear{tCPcB0}
  \be
{\cal E}_{CPcB_\pm} =
-\fr{m^4k^2}{4}\left(\fr{1}{2\sqrt{\lambda_1\lambda_2}}-\fr{\mu^2}{\Lambda_{345-}}\right)
\,.\label{ECPcB0}
\ee

These solutions show that $k$-symmetric vacuum can exist only along the segment
 \be
 |\mu|\le \fr{|\Lambda_{345-}|}{2\sqrt{\lambda_1\lambda_2}}\,,\quad
 \delta=0\label{mudelcondksym}
 \ee
in the $(\mu,\,\delta)$-plane (Fig.~\ref{fig1delta}).

Let us calculate now the difference between extremum energies \eqref{ECPcB0} and \eqref{Apm}.
Remembering that $2\sqrt{\lambda_1\lambda_2}=\Lambda_{345-}+\Lambda_{345+}$, one can obtain easily
 $$
{\cal E}_{CPcB_\pm}-{\cal E}_{CPcA}=\fr{m^2k^2}{4}\;\fr{\left[\Lambda_{345-}(1\pm \mu)\pm \Lambda_{345+}\mu\right]^2}{2\Lambda_{345-}\Lambda_{345+}\sqrt{\lambda_1\lambda_2}}\,.
 $$
From this it follows that the state  B can have a lower energy than the state A, only if $\Lambda_{345-}<0$,
which gives the first condition in \eqref{ksymlamlim} together with useful condition for vacuum state $\mu t_{B+}>\mu t_{B-}>0$.

Now, a direct calculation of masses of the charged and pseudoscalar Higgs bosons for solution B gives
 $$
M_{H^\pm}^2= -v^2\tilde{\Lambda}_{345-}/2\,,\quad M_A^2
= -v^2\Lambda_{3-}/2\,.
 $$
These quantities must be positive at the minimum of the potential, which results in the second and third conditions of \eqref{ksymlamlim}.
The conditions \eqref{ksymlamlim} and \eqref{mudelcondksym} form necessary and sufficient conditions for realization of vacuum $B_\pm$.

For solutions B, it is useful to calculate separately v.e.v.'s for each field $\phi_i$:
\bear{c}
v_{1\pm}^2\equiv \fr{1}{1+t^2}\,v^2=\fr{m^2k^2}{\sqrt{\lambda_1\lambda_2}}(1\mp R),\\[2mm]
v_{2\pm}^2\equiv \fr{t^2}{1+t^2}\,v^2=\fr{m^2}{\sqrt{\lambda_1\lambda_2}}(1\pm R)\,.
 \eear{vi2B0}

{\bf Weak violation of the $\pmb k$-symmetry } (quasi-degenerated case). {\bf Phase transition}.\\
As temperature changes, the quantity $\delta$ goes through zero, see \eqref{straight}.
It is useful to consider the case of  small $\delta$. In this case corrections to $y_i$ and $t$ are easily calculable
from \eqref{Z2tanbeta} considered as solution \eqref{tCPcB0} with a small correction. We do not present here these equations
but show just the corresponding value of extremum energy in the following way:
 \bear{c}
{\cal E}_{CPcB_\pm} =-\fr{m^4k^2}{4}\left[terms\;\;identical\;\;for\;\;{B_\pm}-\right.\\[4mm]
\left.-\mu\tau_{B_\pm}\;\fr{2\mu^2\sqrt{\lambda_1\lambda_2}-\Lambda_{345-}}
{\Lambda_{345-}(\mu^2\sqrt{\lambda_1\lambda_2}+\lambda_{345})}\cdot \delta\right].
 \eear{vacenBpert}
This equation shows that as the curve of physical states passes through $\delta=0$, the phase with the lowest extremum energy
switches from $B_+$ to $B_-$  (or vice versa). That is a phase transition from the phase $B_\pm$
to the phase $B_\mp$ with a jump $v_{1+}\to v_{1-}$, $v_{2+}\to v_{2-}$.
Since the symmetries of these two phases are identical even at $\delta \not =0$, theorems \eqref{2min}
allow for existence of two distinct minima of the potential at small enough $\delta$: the vacuum and a meta-stable state.

\begin{figure*}[t]
\includegraphics[width=0.3\textwidth]{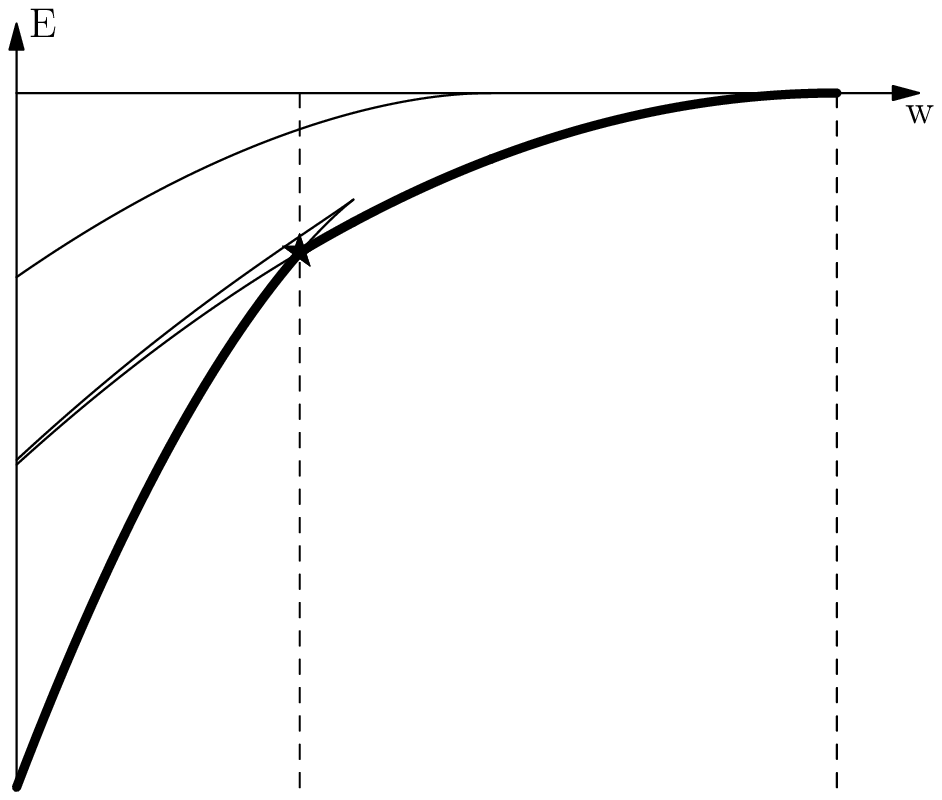}
\hspace{2mm}
\includegraphics[width=0.3\textwidth]{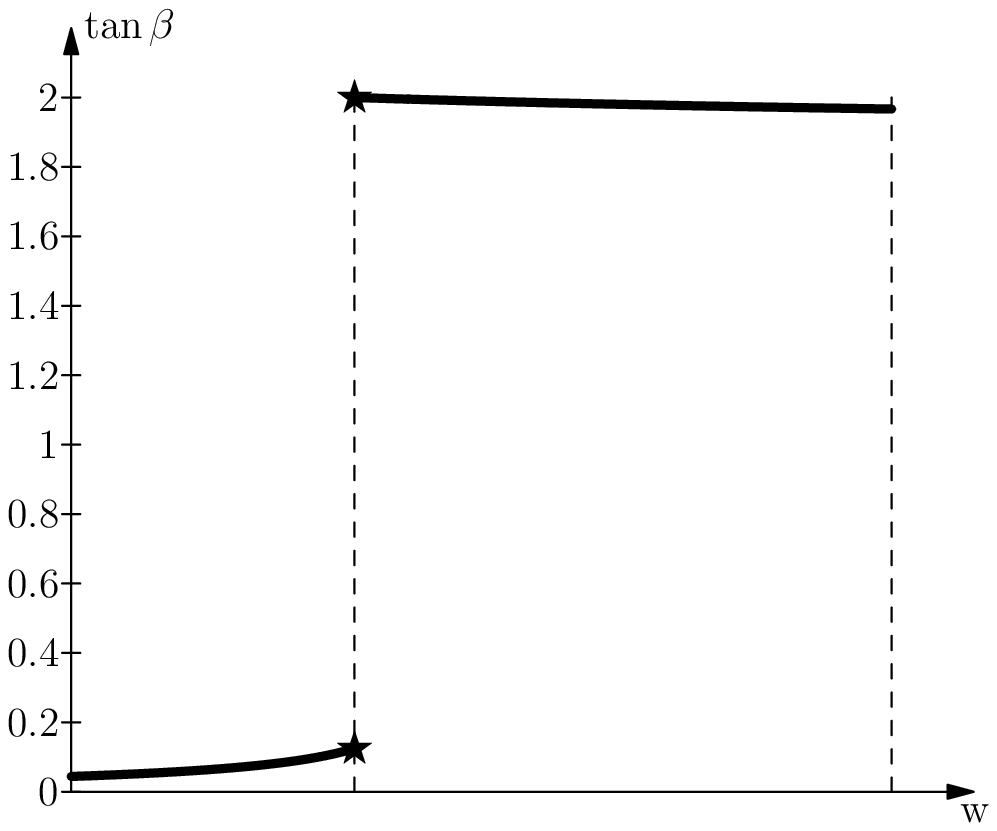}
\hspace{2mm}
\includegraphics[width=0.3\textwidth]{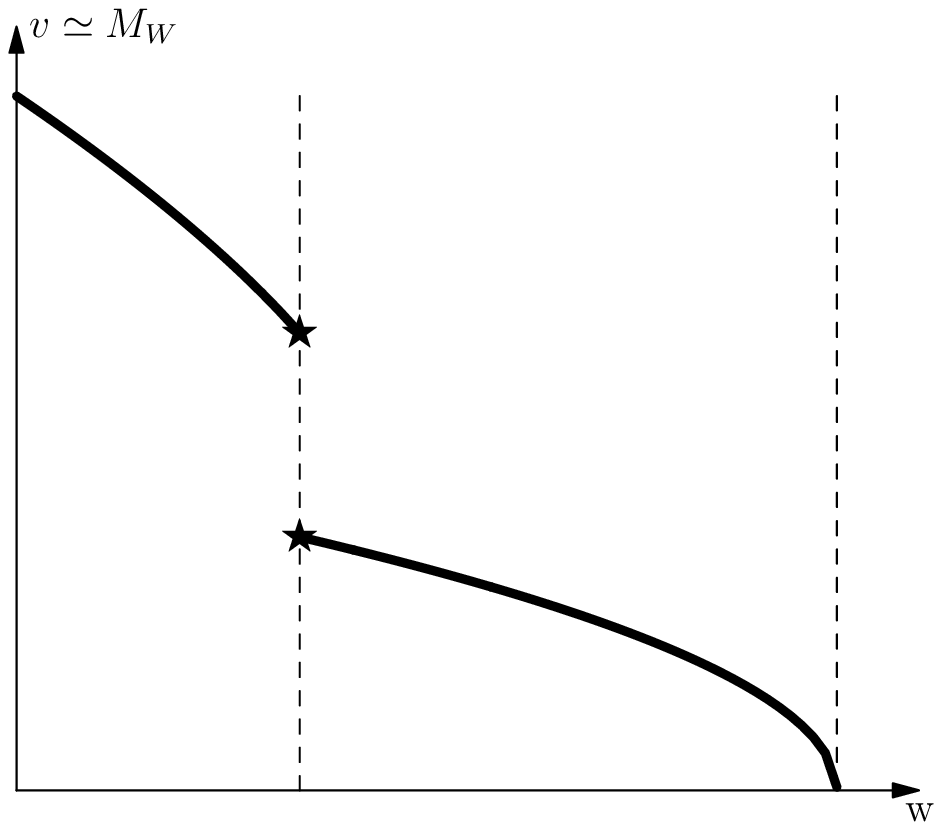}
\begin{center}
	\includegraphics[width=0.3\textwidth]{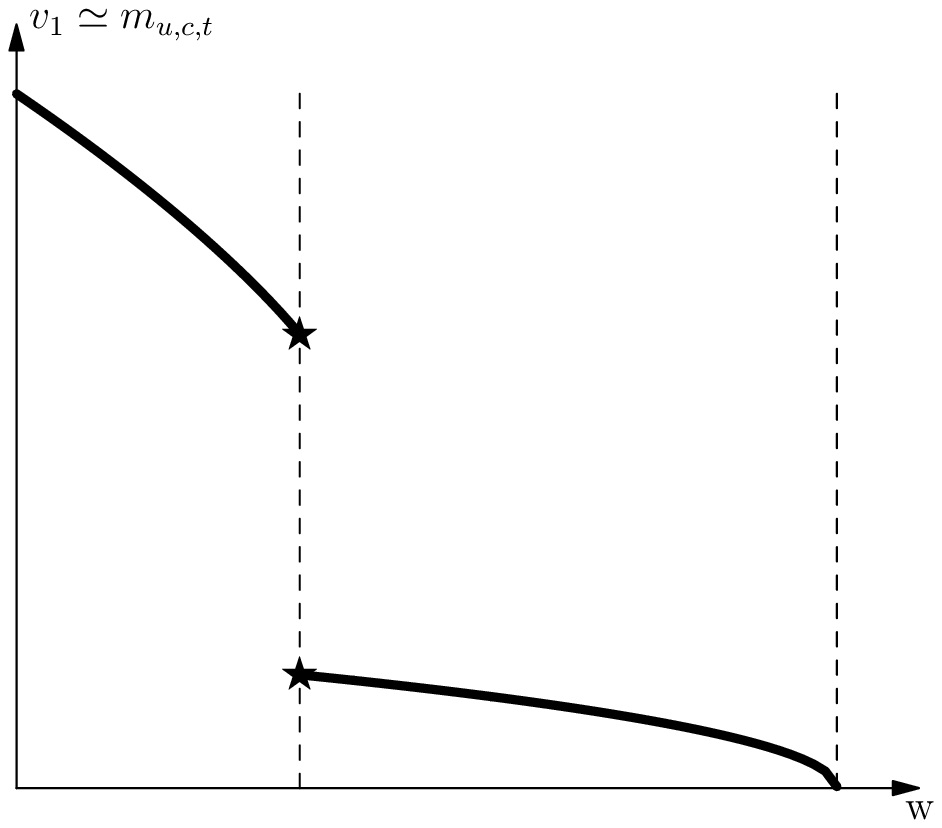}
	\hspace{20mm}
	\includegraphics[width=0.3\textwidth]{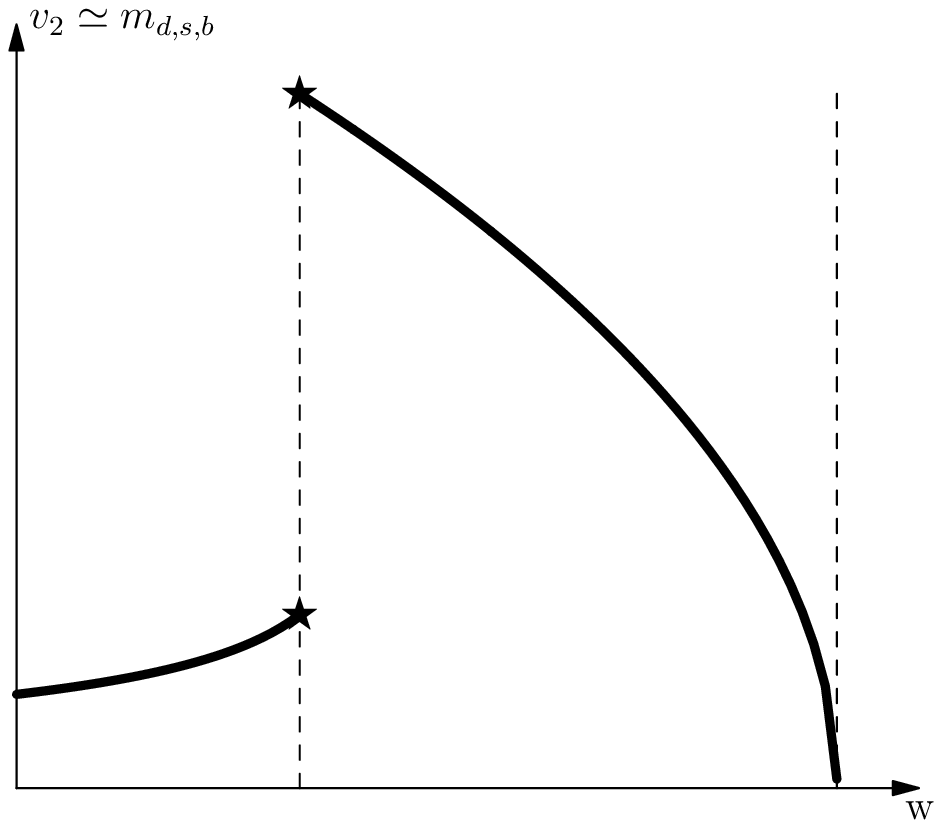}
\end{center}
\caption{\it Evolution of the parameters of the vacuum state for sector II, ray II.
Upper row: Extremum energy, $\tan\beta$, total vacuum expectation value $v$.
Lower row: $v_1$ and $v_2$ shown separately.  }
\label{1stordermainA}
\end{figure*}

The discussed phase transition is a {\bf first order phase transition}. The latent heat of this phase transition
is given by the well known thermodynamical equation (see $c_i$ in sect.~\ref{secdep})
 \bear{c}
 Q_{-\to +}=\left. T\fr{\pa{\cal E}_+}{\pa T}-T\fr{\pa{\cal E}_-}{\pa T}\right|_{\delta\to 0}=\\[3mm]
 =\fr{m^4k^2}{2}\;\left(\fr{2\mu^2\sqrt{\lambda_1\lambda_2}-\Lambda_{345-}}
{\mu^2\sqrt{\lambda_1\lambda_2}+\lambda_{345}}\right)\;
\fr{R}{\sqrt{\lambda_1\lambda_2}}\,(c_1-c_2)w\,.
\eear{latheat}

{\bf The case with non small $\delta$} can be studied by directly solving eq.~\eqref{Z2tanbeta}.
These solutions are direct continuation of above quasi-degenerated case.

\bu {\bf The $\pmb{ (\mu,\,\delta)}$ plane}. Fig.~\ref{fig1delta} represents the first sheet of the $(\mu,\,\delta)$ plane for
some specific sets of parameters $\lambda_i$ from sector II.
Several  possible ``zero point'' states (i.e. different choices of zero-temperature $\mu$ and $\delta$) are shown.
The first order phase transition segment, described by eq.~\eqref{mudelcondksym}, is marked by thick line.
Possible evolution paths of physical states are presented by rays, arrows indicating the direction of temperature growth.

Small dots on these rays labeled 1,2, and 3 correspond to possible ``zero points''.
The shaded area covers all ``zero points'' that would lead to crossing of the thick segment at some temperature, which would force
the system to go through a first order phase transition (e.g. point 2). In this case the phase evolution during cooling down of the Universe was\\
\cl{{\it EWs $\to$ CPc1 $\to$ CPc2,}}
that is, the Universe experienced two phase transitions: EWSB, of the second order, and CPc1 $\to$ CPc2, of the first order).

If today's state corresponds to the ``zero point'' 3 \ \ lying on ray II, or if it is located anywhere on ray I, then the Universe
has experienced only one standard EWSB phase transition during its
cooling down\fn{Note that at $k=1$ we have $c_1=c_2\Rightarrow {\cal {P}}_0=(0,\,0)$. The ray of physical states never crosses
the line segment of the first order phase transitions $\delta=0$, so, we have no phase transition.}:\\
\ \cl{\it EWs $\to$ CPc.}
This case was described in the previous subsection.

\begin{figure*}[t]
\includegraphics[width=0.3\textwidth]{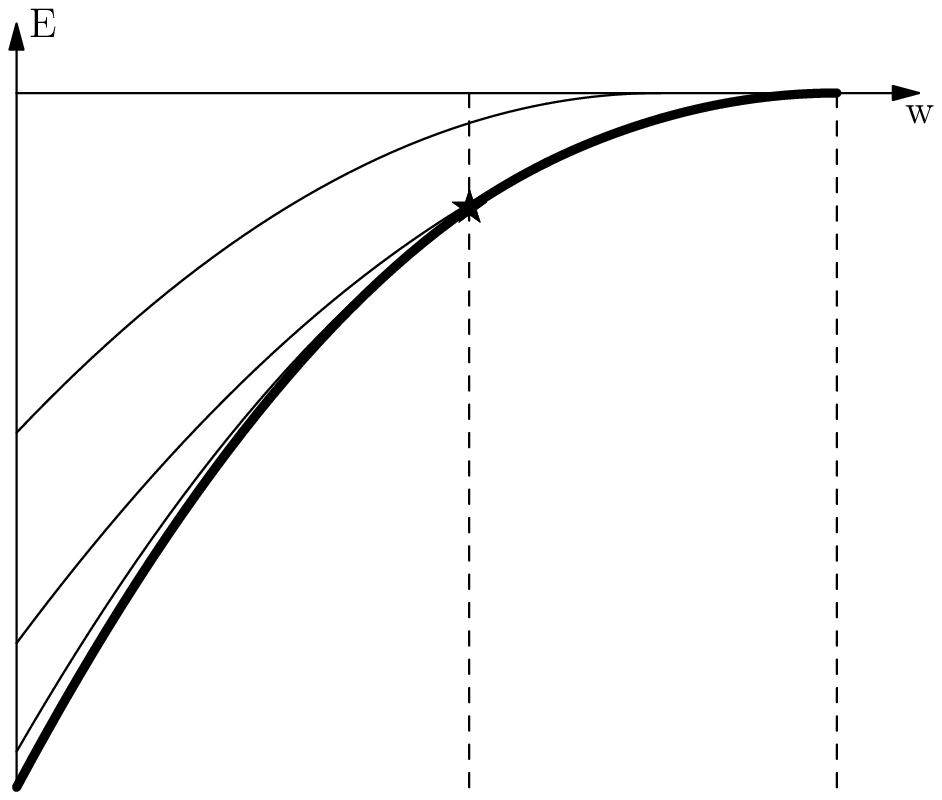}
\hspace{2mm}
\includegraphics[width=0.3\textwidth]{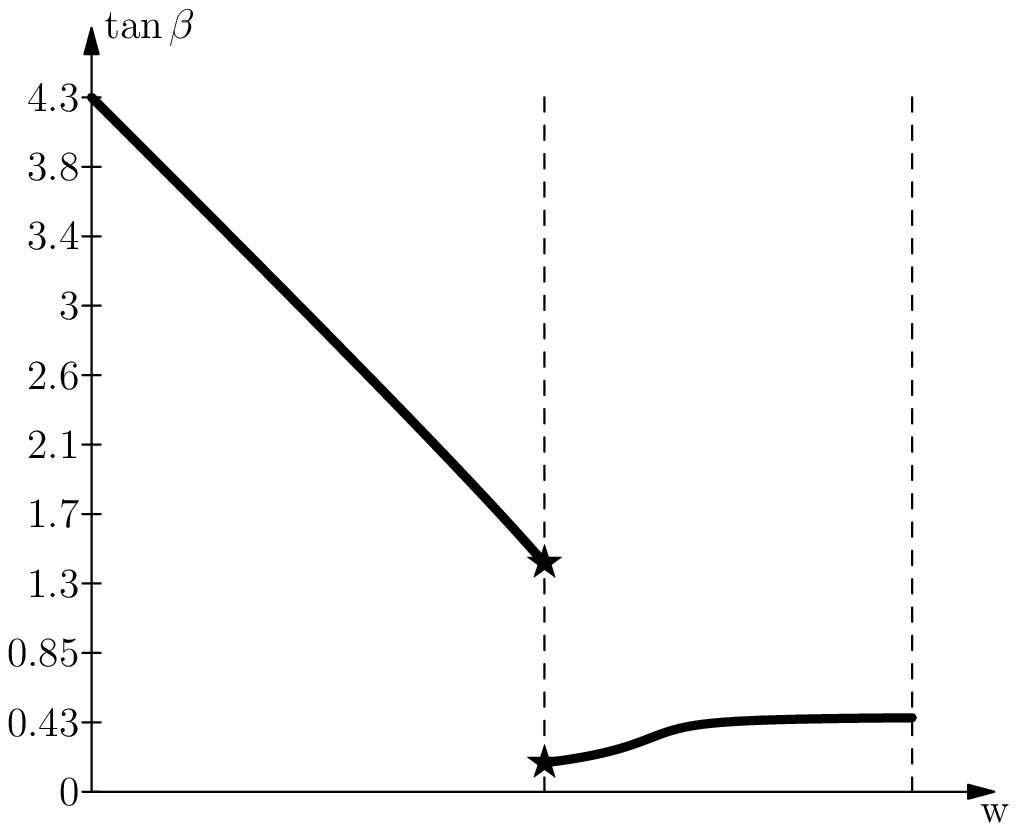}
\hspace{2mm}
\includegraphics[width=0.3\textwidth]{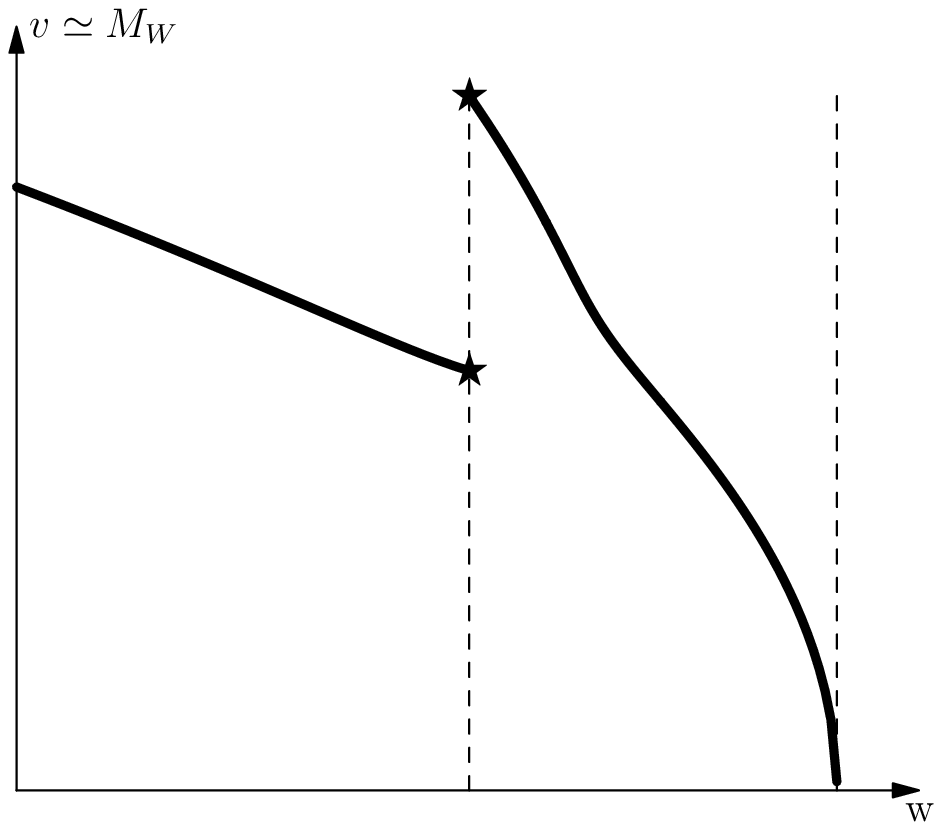}
\begin{center}
	\includegraphics[width=0.3\textwidth]{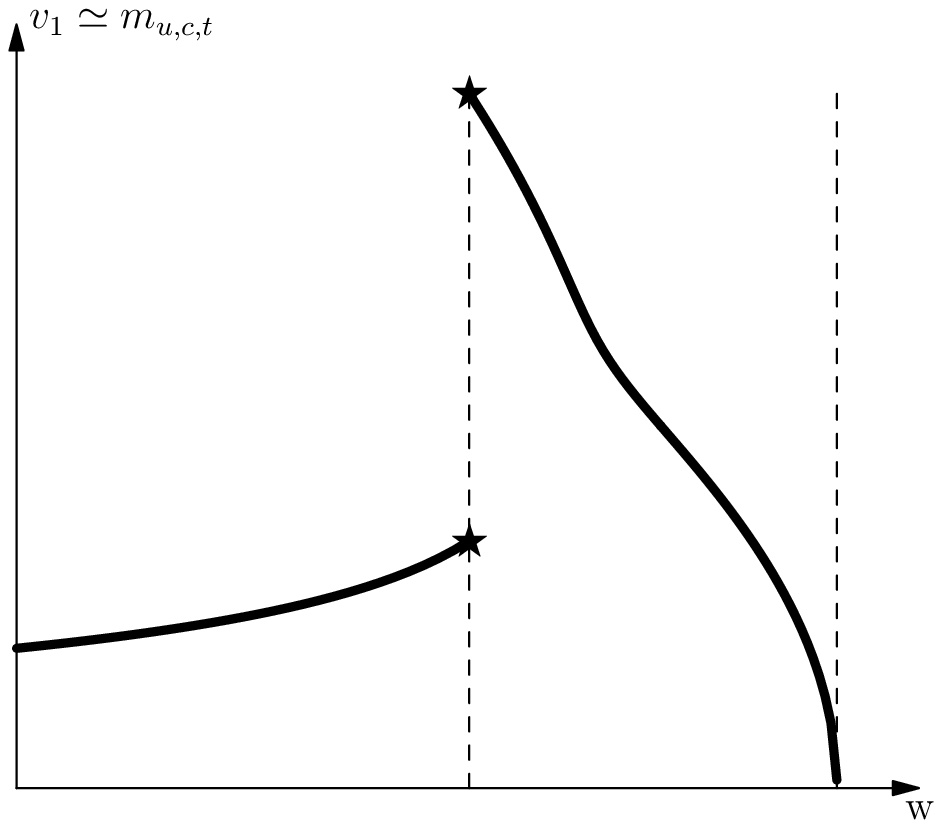}
	\hspace{20mm}
	\includegraphics[width=0.3\textwidth]{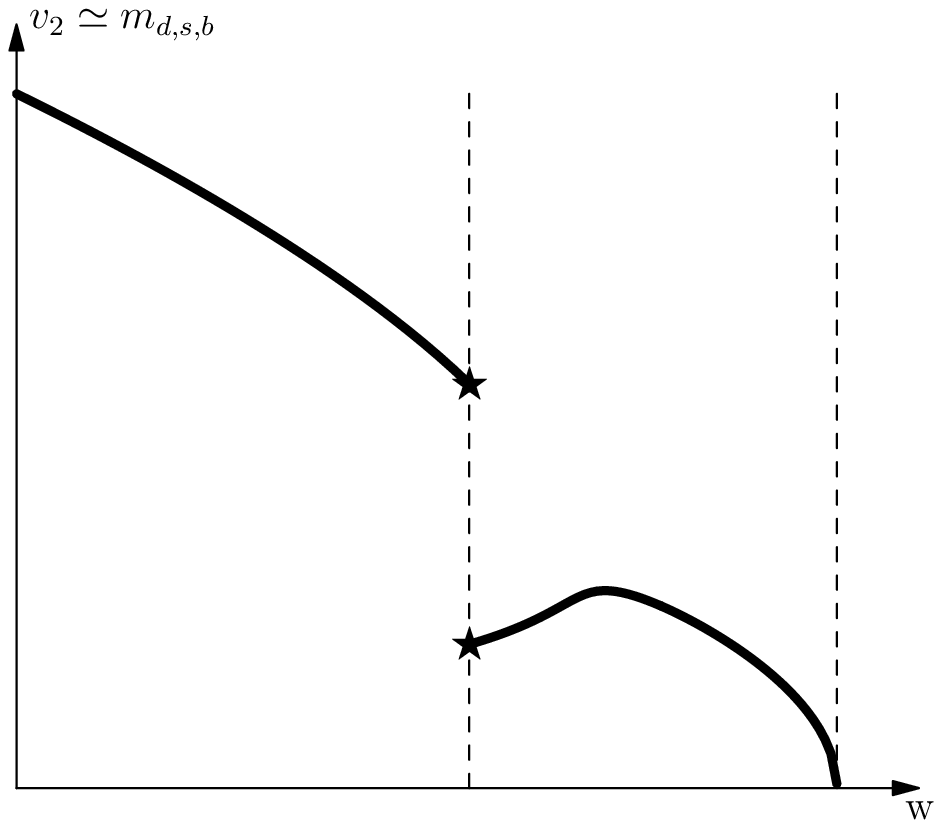}
\end{center}
\caption{\it The same as in Fig.~\ref{1stordermainA} but for another set of $\lambda_i$.
}
\label{1stordermainB}
\end{figure*}

\bu {\bf Evolution of the parameters of vacuum state}.
A typical evolution of the parameters of the vacuum state for ray I has already been shown above in
Figs.~\ref{1CPcfigA}, Figs.~\ref{1CPcfigB}.
Figs.~\ref{1stordermainA}, \ref{1stordermainB} represent evolution of the parameters of the vacuum for ray II starting from the ``zero point'' 2.
The same graphs can be also used to track the evolution of the same ray, but starting from another ``zero point'' 3.
In all these Figures we show evolution of the parameters of the vacuum state as functions of the parameter $w$ \eqref{Tempdep},
the temperature increasing from left to right.
The evolution of the physical parameters of the true vacuum state is shown everywhere by thick lines,
while the thin lines correspond to the other extrema and are shown just for comparison.
The phase transition points are indicated by small stars.

In all pictures the upper left plots represent thermal evolution of the extrema energies shown with thin lines.
The vacuum state is drawn with thick line.
The upper middle plots show the evolution of $\tan\beta=v_2/v_1$, while the upper right plots gives
the temperature dependence of the main v.e.v. value $v=\sqrt{v_1^2+v_2^2}=M_W/g$.
The lower row shows the two v.e.v.'s $v_1$ and $v_2$ separately.
Within Model II for the Yukawa sector $v_1\propto m_u \,(m_t)$, $v_2\propto m_d\,(m_b)$.

The first order nature of the phase transition is evident on these plots,
as the quantities $v$, $v_i$ and $\tan\beta$ jump at the transition point.
These changes can be large in their magnitude.
One can see that for the different sets of $\lambda_i$, distinct variants of thermal evolution can be realized.
For the cases shown in Fig.~\ref{1stordermainA} (Fig.~\ref{1stordermainB}) the following changes take place at the phase transition, respectively:\\
\bu $\tan\beta$ jumps up (down), \\
\bu main v.e.v. $v$ as well as the masses of the gauge bosons $M_W$, $M_Z$ jump down (up), so that the today's value of $v$
can be either larger or smaller than before the phase transition.
Similar behavior can be demonstrated by $v_i$ separately.
In particular, Fig.~\ref{1stordermainB} displays a peculiar $M_W\propto v$ behavior as temperature increases:
at first it monotonically decreases, then jumps up at the first order phase transition, and then
goes down again and turns zero at the point of EWSB.

\begin{figure*}[t]
\begin{center}\includegraphics[height=3cm,width=0.24\textwidth]{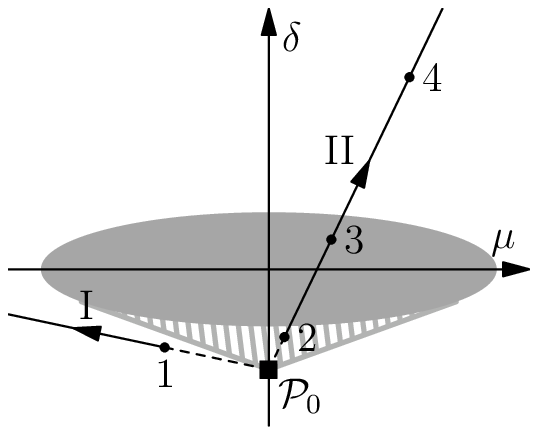}
\hspace{6mm}
\includegraphics[height=3cm,width=0.24\textwidth]{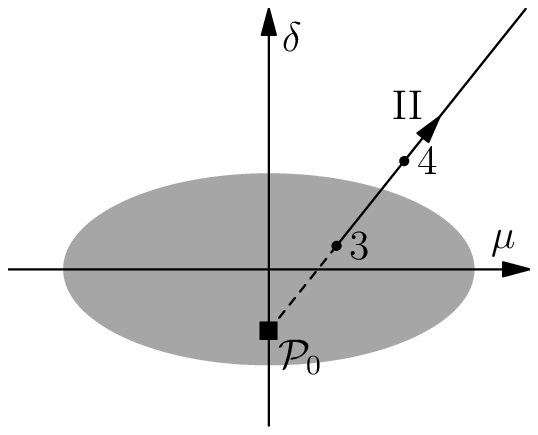}
\end{center}
\vspace{-5mm}
\caption{\it The $(\mu,\,\delta)$ plane for sector III.}
\label{delmuplane}
\end{figure*}

\begin{figure*}[t]
\includegraphics[height=4cm,width=0.3\textwidth]{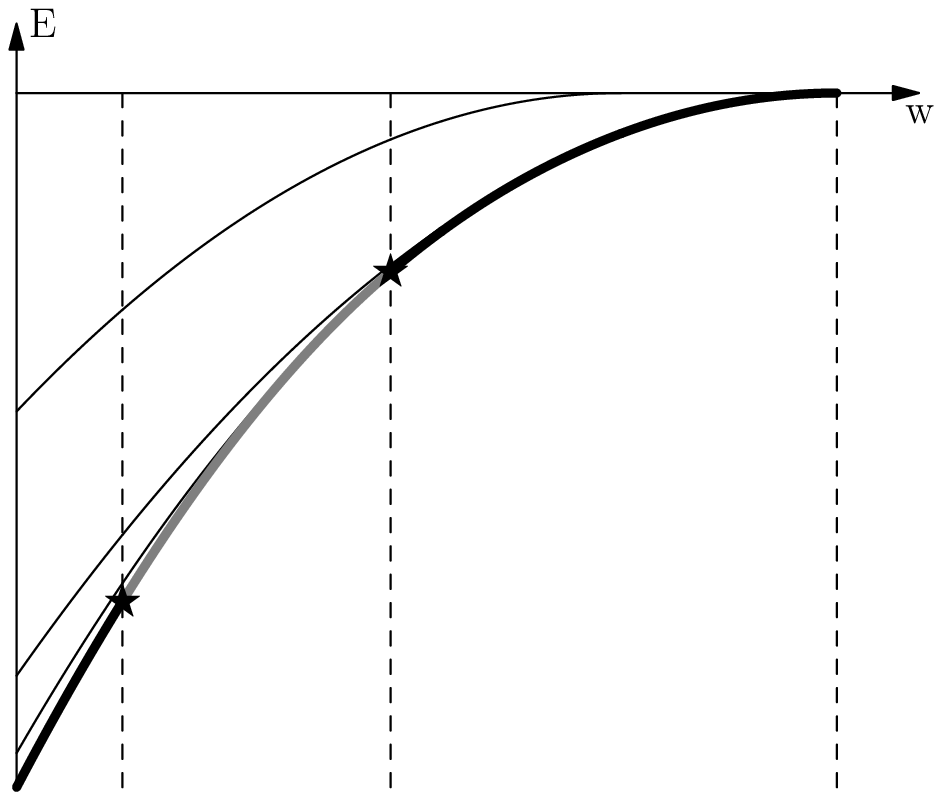}
\hspace{2mm}
\includegraphics[height=4cm,width=0.3\textwidth]{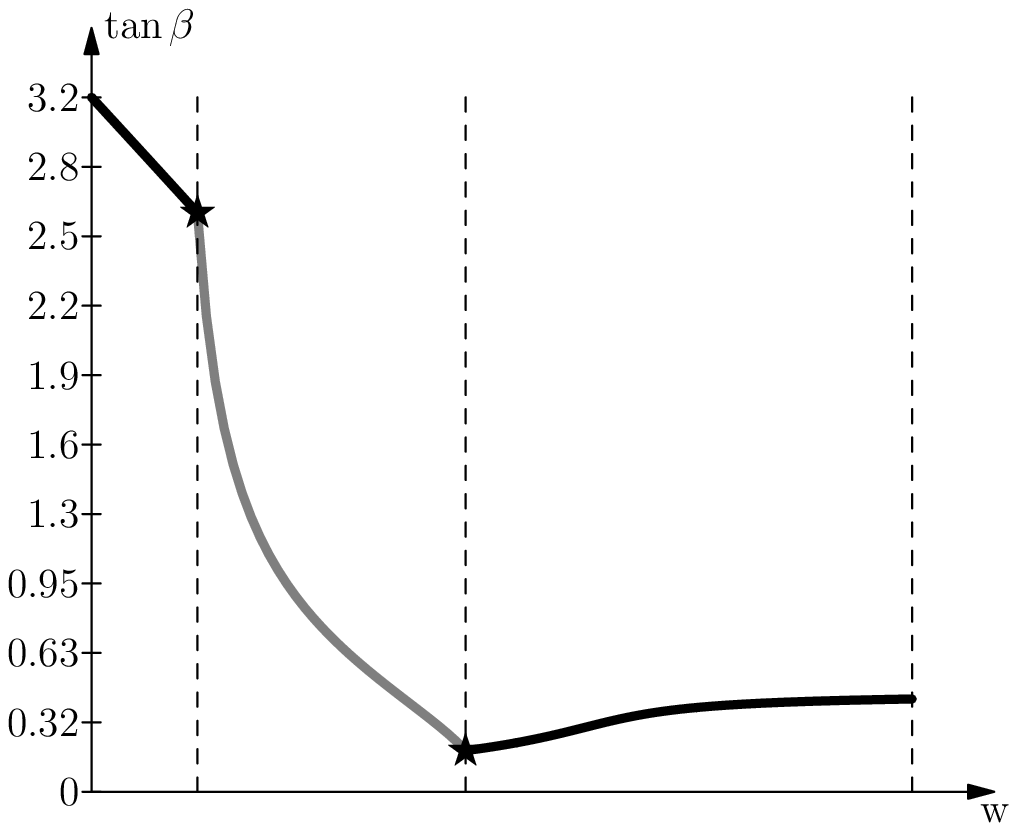}
\hspace{2mm}
\includegraphics[height=4cm,width=0.3\textwidth]{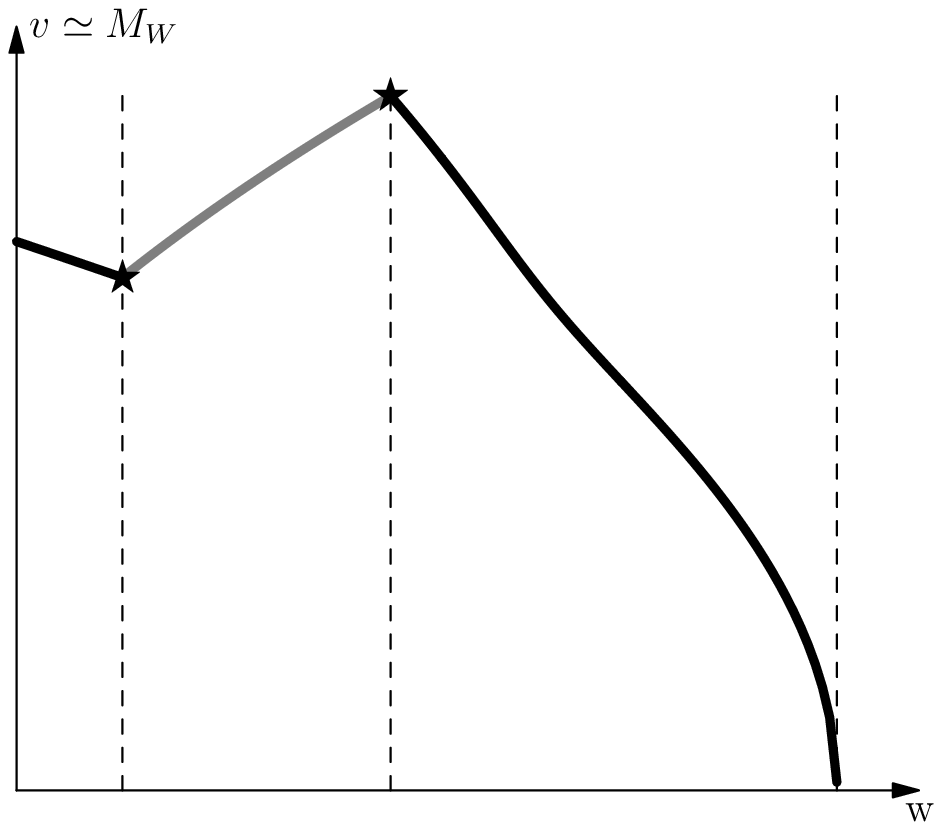}
\\[5mm]
\includegraphics[height=4cm,width=0.3\textwidth]{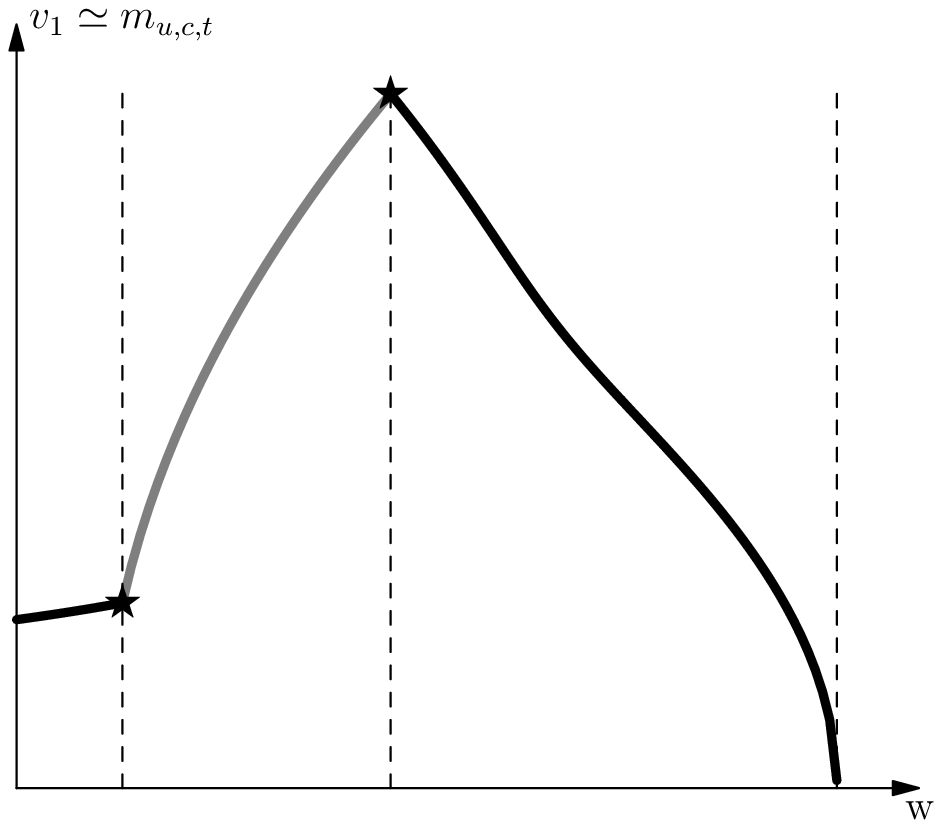}
\hspace{2mm}
\includegraphics[height=4cm,width=0.3\textwidth]{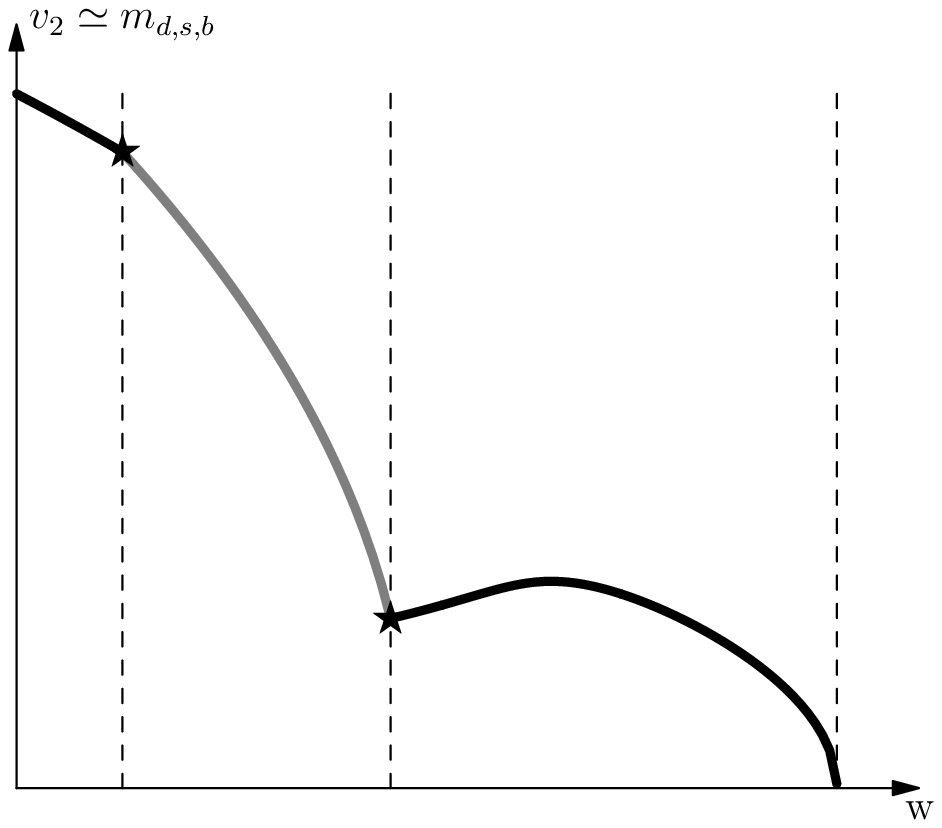}
\hspace{2mm}
\includegraphics[height=4cm,width=0.3\textwidth]{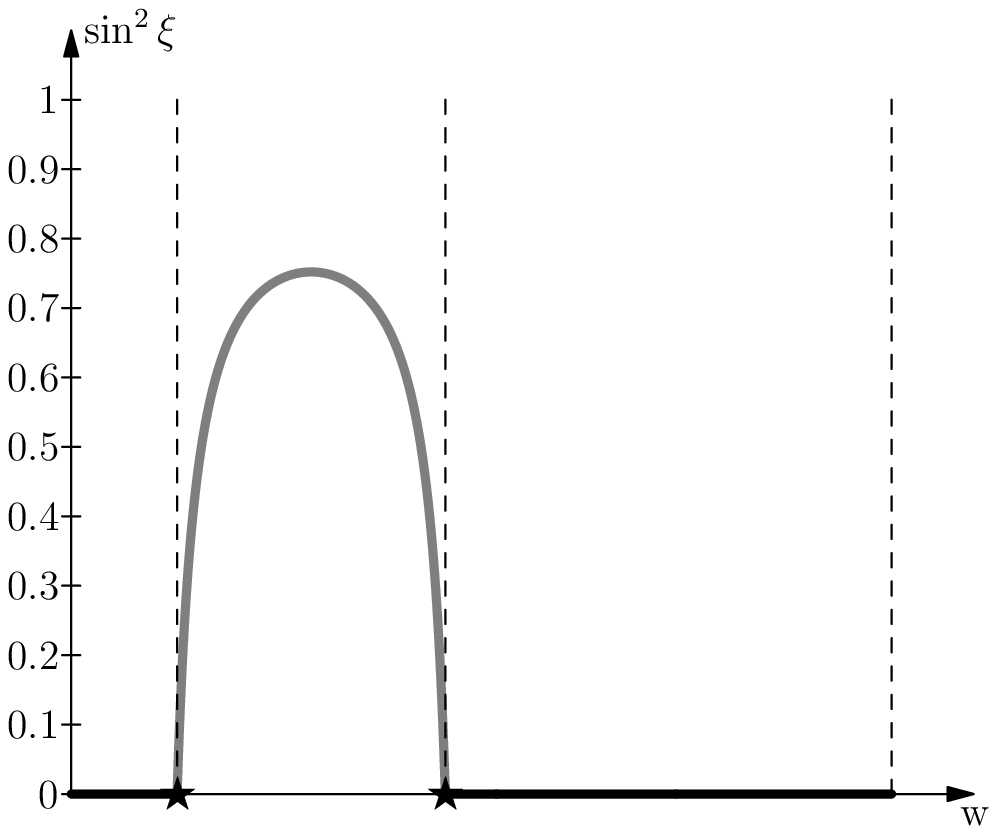}
\caption{\it
Evolution of the parameters of the vacuum state though the double $CP$-violating and
$CP$-restoring phase transition.
Upper row: Vacuum energy evolution (left), the insert zooms at the region between the two
phase transitions; $\tan\beta$ (center); $v$ (right). Lower row:
$v_1$ (left); $v_2$ (center); $\sin^2\xi$ (right).  }
\label{sCPvmainA}
\end{figure*}

\subsection{Sector III. Spontaneously CP violating (sCPv) minimum}

The spontaneously CP violating vacuum can exist only in {\bf sector III}, which is defined by the inequalities
\bear{c}
\lambda_5>0, \quad \lambda_5>\lambda_4,\quad
\tilde{\Lambda}_{345-}>0\,.
\eear{scPvlam}
We again will explain the origin of these limitations in this section.

Since the potential is $CP$ symmetric, the sCPv extremum is doubly degenerated in the sign of the $CP$ violating phase $\xi$.
Because of the property \eqref{2min}, if a sCPv minimum exists, it is necessarily the global one, i.e. the vacuum (for more details see \cite{GK07}).

To describe a sCPv extremum, it is useful to express the potential \eqref{Z2potential} via quantities $y_1$, $y_2$ and $\xi$.
This gives a second-degree polynomial in $\cos\xi$. Therefore
the extremum condition can be solved first for $\cos\xi$. It has two solutions, the CPc solution $\sin\xi=0$ and sCPv solution with $\xi\neq 0$.
We discuss here only the sCPv solution.
The coefficient in front of the $\cos^2\xi$ term in the potential is proportional to $\lambda_5$,
so if we want sCPv to be the minimum, we must require than $\lambda_5>0$, which yields the first inequality of \eqref{scPvlam}.
Besides, a direct calculation of the charged Higgs boson mass for the sCPv phase, \cite{GK07},
gives $M_{H^\pm}^2=(\lambda_5-\lambda_4)v^2/2$, whose positivity leads to the second inequality in
\eqref{scPvlam}.

Inserting $\cos\xi$ obtained back into the extremum condition yields a second-degree polynomial in  $y_{1,\,2}$.
The minimum conditions for $y_i$ take form of two simultaneous linear equations,
which have unambiguous solutions for any values of the parameters (without assumption on $Z_2$ symmetry). Finally
  \bes\label{yiSPCPcoftZ}
  \bea
 &y_1\!=\fr{k^2m^2}{2}\left[\fr{1}{\tilde{\Lambda}_{345+}}
 -\fr{\delta}{\tilde{\Lambda}_{345-}}\right],\\[2mm]
 &y_2\!=\fr{m^2}{2}\left[\fr{1}{\tilde{\Lambda}_{345+}}
 +\fr{\delta}{\tilde{\Lambda}_{345-}}\right],&\label{ysCPv}\\[2mm]
&\cos\xi\equiv c_0=\fr{\mu km^2}{4\lambda_5\sqrt{y_1y_2}}\,;&\label{xisCPV}\\[3mm]
 &{\cal E}_{sCPv}=-\fr{m^4k^2}{4}
 \left[\fr{1}{\tilde{\Lambda}_{345+}}
 +\fr{\delta^2}{\tilde{\Lambda}_{345-}}+\fr{\mu^2}{2\lambda_5}\right]
 \,.&\label{EsCPv}
 \eea\ees
Equation \eqref{xisCPV} has two solutions
 \be
\xi_+=\arccos c_0\,,\qquad  \xi_-=-\,\xi_+\,.\label{2xi}
 \ee
The  vacuum energy ${\cal E}_{sCPv}$ \eqref{EsCPv} is the same for both these solutions:
the vacuum state is doubly degenerate in the sign of the $CP$ violated phase $\xi$.

Simple algebra allows one to rewrite the condition $\cos^2\xi\le 1$ as the condition that the sCPv state
lies only inside a specific ellipse in the $(\mu,\,\delta)$ plane:
\bear{c}
\fr{\mu^2}{b_1^2}+\fr{\delta^2}{b_2^2}=1, \quad b_1=\fr{2\lambda_5}
{\tilde{\Lambda}_{345+}}, \quad b_2=\fr{\tilde{\Lambda}_{345-}}{\tilde{\Lambda}_{345+}}.
 \eear{sCPvlim}
A geometrical analysis shows that the necessary conditions for the sCPv vacuum contains the condition $\tilde{\Lambda}_{345-}>0$,
which is the third inequality in \eqref{scPvlam}, \cite{Ivan1}.

\begin{figure*}[t]
\includegraphics[width=0.3\textwidth]{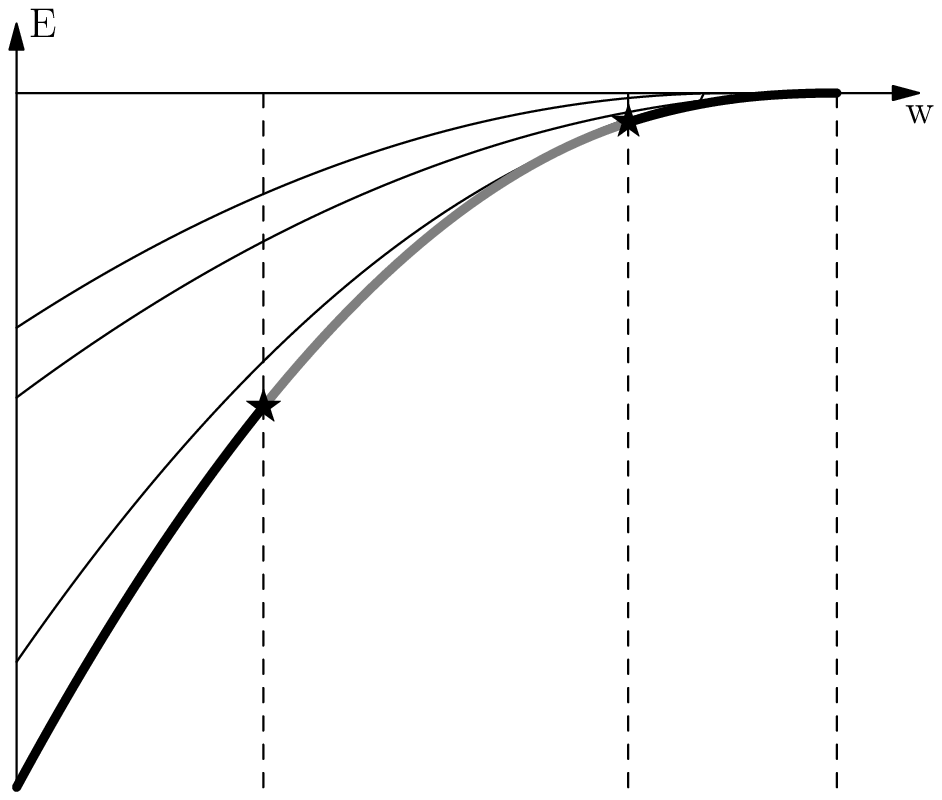}
\hspace{2mm}
\includegraphics[width=0.3\textwidth]{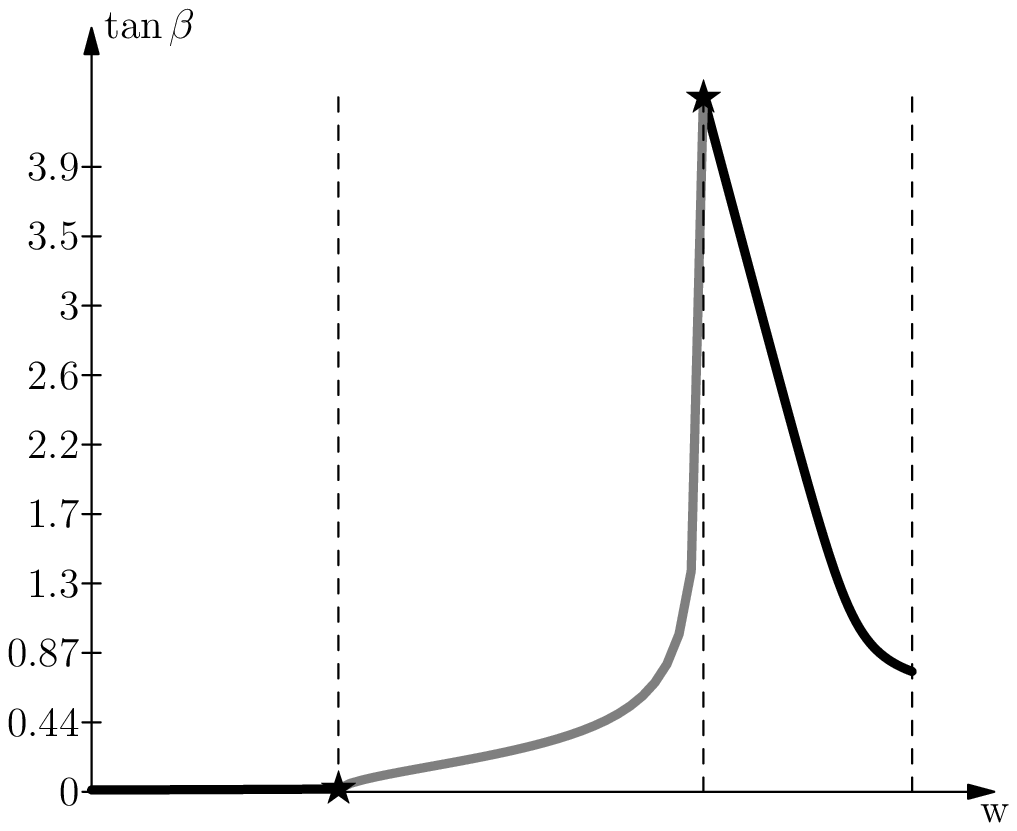}
\hspace{2mm}
\includegraphics[width=0.3\textwidth]{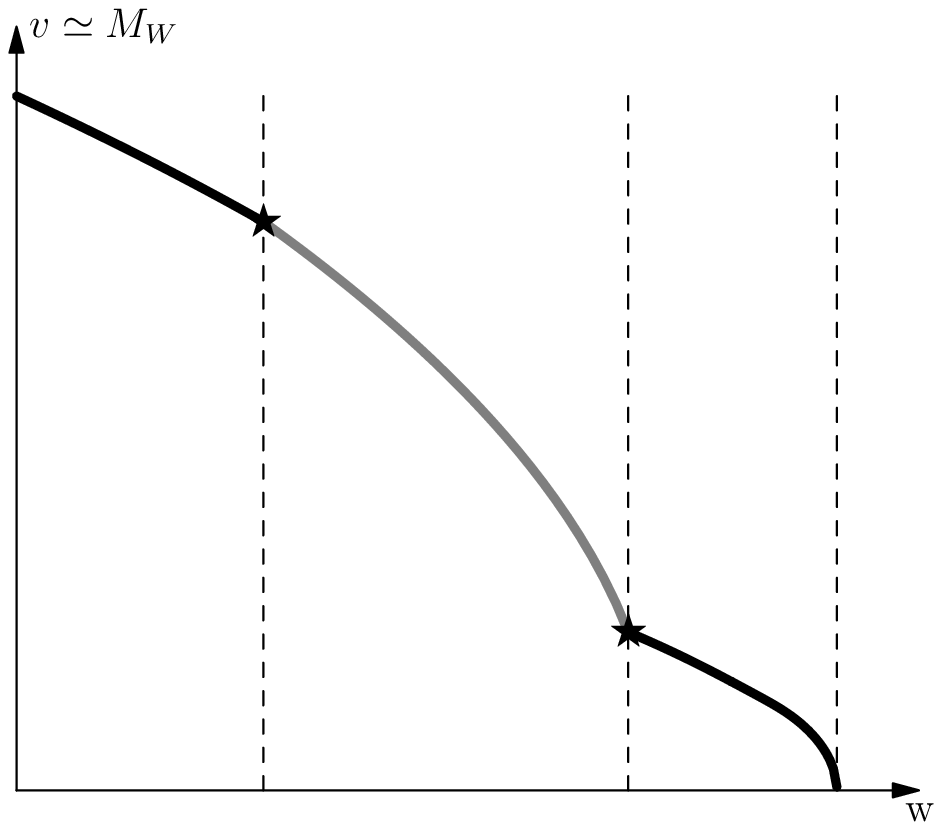}\\[5mm]
\includegraphics[width=0.3\textwidth]{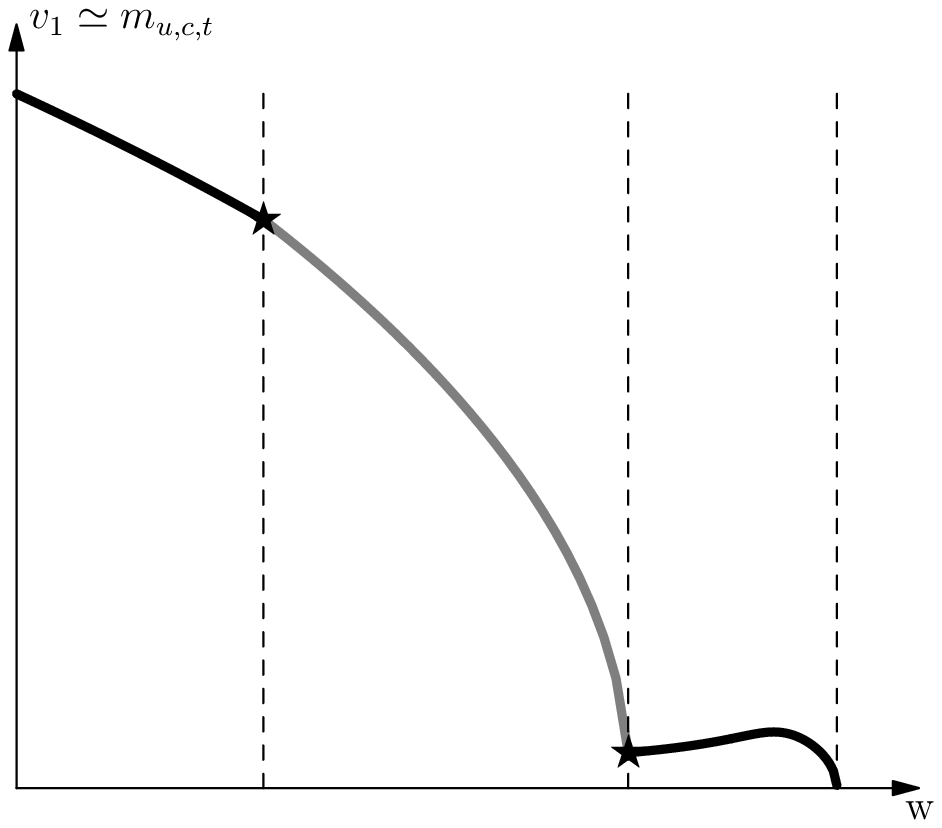}
\hspace{2mm}
\includegraphics[width=0.3\textwidth]{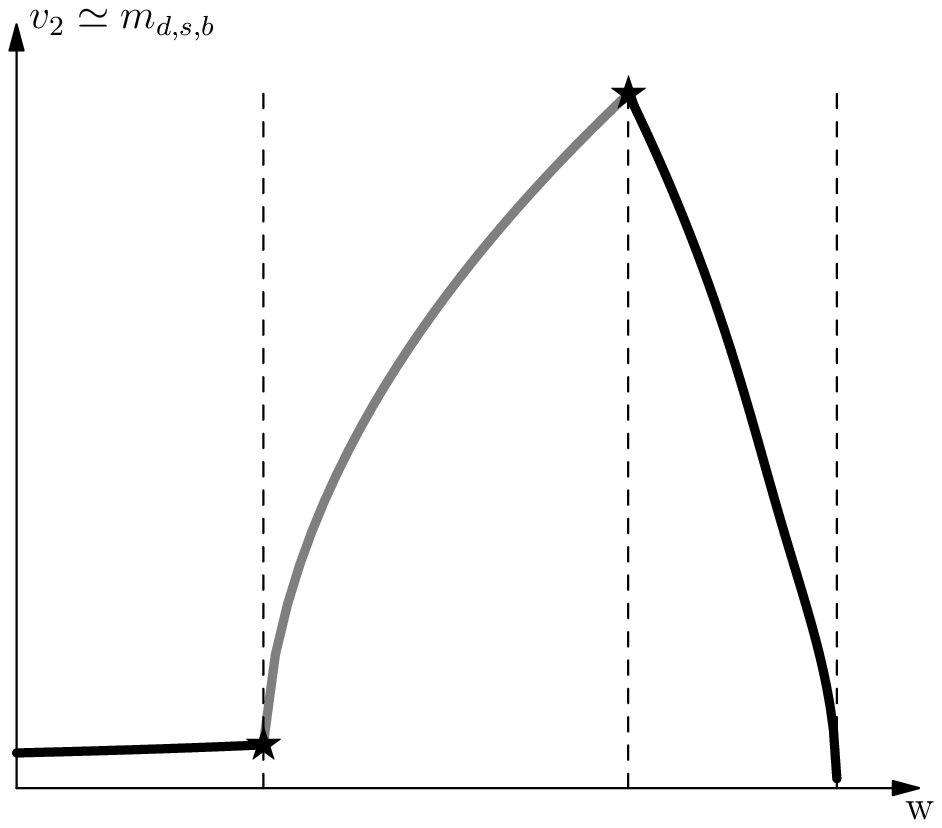}
\includegraphics[width=0.3\textwidth]{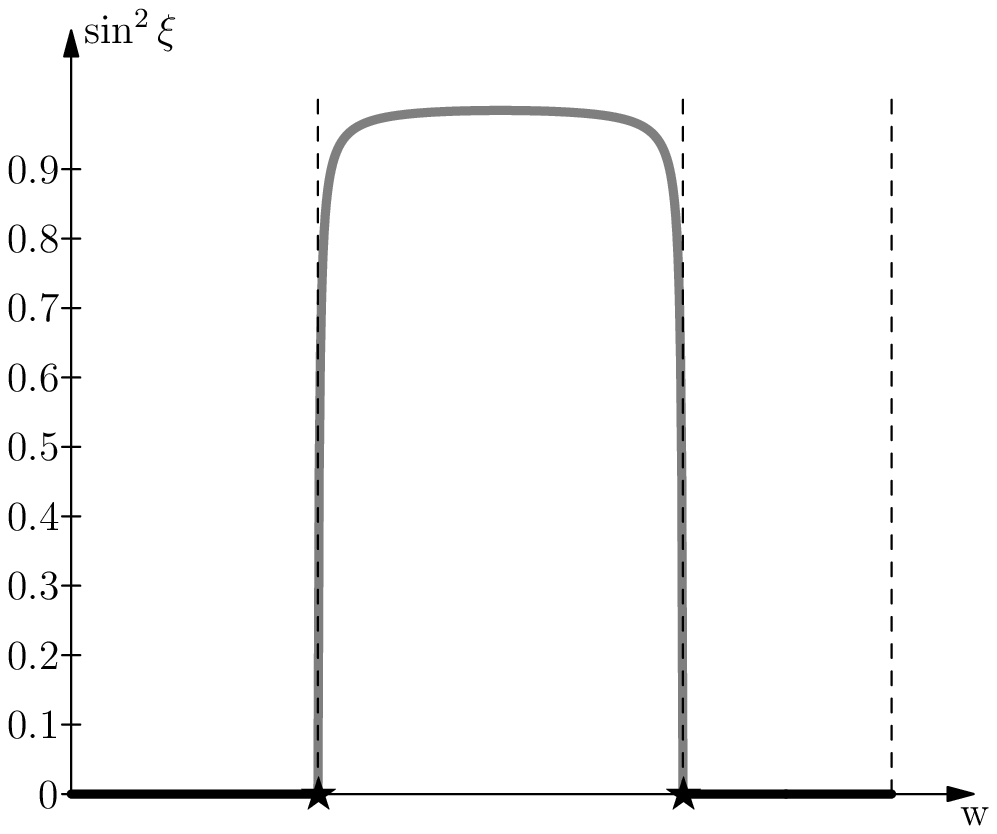}
\caption{\it The same as in Fig.~\ref{sCPvmainA} for another set of parameters.}
\label{sCPvmainB}
\end{figure*}

\bu {\bf The $\pmb{ (\mu,\,\delta)}$ plane}. Fig.~\ref{delmuplane} shows the first sheet of the $(\mu,\,\delta)$ plane for different sets of $\lambda_i$ from sector III. The states of the sCPv vacuum are located within the gray ellipse, described by eq.~\eqref{sCPvlim}.
The left plot represents a case when the point ${\cal P}_0$ lies outside the ellipse, while the right plot refers to a situation when the point ${\cal P}_0$ is inside the ellipse.

Possible evolution of physical states is again presented here by rays, with arrows indicating the direction of the temperature growth. The hatched area covers all the modern values of $(\mu,\,\delta)$, such as point 2 on ray II in left plot, for which thermal evolution
in the past went through the ellipse of sCPv vacua with two phase transitions. For these states phase evolution during cooling down of the Universe was\\[2mm]
 \cl{{\it EWs $\to$ CPc $\to$ sCPv $\to$ CPc}}\\ \cl{(three phase transitions of the second order).} \\[2mm]

Today's states such as 3 describe a sCPv vacuum (for both plots in Fig.~\ref{delmuplane}).
For these states the phase evolution during cooling down was\\[2mm]
 \cl{{\it EWs $\to$ CPc $\to$ sCPv }}\\ \cl{(two phase transitions of the second order).} \\[2mm]

If today's state corresponds to the ``zero point'' 4 on ray II, or if it is located anywhere on ray I,
then the Universe during its thermal evolution had just
one standard EWSB phase transition, the case already described above.\\

\bu {\bf Evolution of parameters of the vacuum state}.
Thermal evolution of the parameters of the vacuum for ray I is the same as in Figs.~\ref{1CPcfigA},~\ref{1CPcfigB}.
We discuss here the evolution of the parameters of the vacuum state along ray II of Fig.~\ref{delmuplane},
which is shown in Figs.~\ref{sCPvmainA},~\ref{sCPvmainB} and which reflects a series of second order phase transitions
CPc1$\to$sCPv$\to$CPc2 (the phases CPc1 and CPc2 can be either different or identical).
Notation and description of these plots are the same as for the previous case Fig.~\ref{1stordermainA}.
An essentially new picture, shown last in the lower row, represents the behavior of the order parameter for sCPv case, $\sin^2\xi$,
which is, of course, non-zero only for the sCPv phase.

It is interesting to note that for the set of $\lambda_i$ chosen in Fig.~\ref{sCPvmainA}, the energy difference between
the vacuum and the second deepest extremum is small and practically invisible on the main graph.
It makes fluctuations from the vacuum sCPv state into the low-lying CPc states (saddle points) not that much suppressed.

The order parameter $\sin^2\xi$ exhibits a similar behavior in both cases.
Thermal evolution of the other parameters makes it evident that we deal with second order phase transitions.
The curves are continuous but their slopes experience jumps at the transition points (indicated by small stars),
similar to e.g. susceptibilities in condensed matter physics.

All other features of evolution of the physical parameters can be different as in previous cases.
V.e.v.'s can either decrease monotonically or increase at some temperature.
In particular, for the case of Fig.~\ref{sCPvmainA}  with going to the past in the CPc phase
the mass $M_W\propto v$ decreases (together with mass scale parameter $m^2(T)$)
but in the sCPv phase it grows strong with increase  of temperature.
After second phase transition all masses become decrease monotonically to the past,
starting from new level, heavier than modern (for $M_W$ and $m_u$).

\begin{figure*}[ht]
\begin{center}\includegraphics[height=3cm,width=0.24\textwidth]{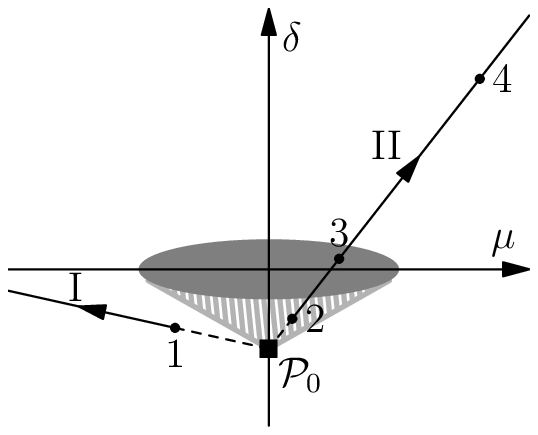}
\hspace{6mm}
\includegraphics[height=3cm,width=0.24\textwidth]{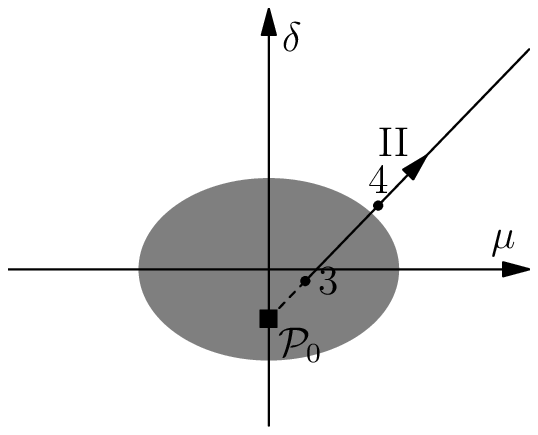}
\end{center}
\caption{\it The $(\mu,\,\delta)$ plane for sector IV}
\label{charged_dmplane}
\end{figure*}

\subsection{Sector IV. The charge-breaking vacuum (with  $\pmb {u\neq 0}$)}\label{seccharge}

The charge-breaking vacuum can exist only in {\bf sector IV}, which is defined by the following inequalities (fore more details see \cite{GK07}):
\be
\vspace{-1mm}
\lambda_4 \pm \lambda_5>0\,,\quad \Lambda_{3-}>0\,.\label{Lamch}
\vspace{-2mm}
\ee
In a charged vacuum, it is not possible  to split the Higgs boson and the gauge boson mass matrices
into the neutral and charged sectors, the interaction of gauge bosons with fermions will not preserve electric charge, photon becomes massive, etc.
\cite{Barroso}.
This is the reason why such vacuum is called charge-breaking.
Certainly, this phase cannot describe today's Universe, however the vacuum could have evolved through this phase in the past.

In the problem of finding the charged extremum one can treat variables $y_i$ as independent,
so that the position of this extremum  is given by the system of linear equations, written in variables $y_i$,
$\pa V/\pa y_i=0\;\Rightarrow\; \Lambda_{ij}y_{j,ch}={\cal M}_i$.
 If $\Lambda_{ij}$ is not singular, a solution to this system always exists and is unique (but it might violate conditions \eqref{Zcond}).

Decomposition \eqref{decomp} shows that if such an extremum realizes the minimum of the potential, this minimum is the global one (vacuum).
This decomposition shows that the charged extremum can be minimum if only $V_4(x_i)>0$
for {\bf all} classical $x_i$ regardless of conditions \eqref{Zcond} ({\it the
necessary condition}).
In particular,  for the potential \eqref{Z2potential} it means that the conditions \eqref{positivsoft}
are to be accompanied with inequalities \eqref{Lamch}.
For this potential the extremization problem
is solved easily, and after simple algebra the solutions can be written \cite{GK07} in form
 \bear{c}
y_1=\fr{m^2k^2}{2}\left(\fr{1}{\Lambda_{3+}}
-\fr{\delta}{\Lambda_{3-}}\right),\\[2mm]
y_2=\fr{m^2}{2}\left(\fr{1}{\Lambda_{3+}}
+\fr{\delta}{\Lambda_{3-}}\right),\\[3mm]
y_3=\fr{m^ 2k\mu
}{2(\lambda_{4}+ \lambda_{5}) }\,;\\[4mm]
{\cal E}_{ch}^{ext}\!=\!-\fr{m^4k^2}{4}\left[\fr{1}{\Lambda_{3+}}
+\fr{\delta^2}{\Lambda_{3-}}
+\fr{\mu^2}
{\lambda_{4} +\lambda_{5}}\right]\,.
  \eear{chargevacyi}
Simple algebra shows that condition $Z>0$ \eqref{Zcond}, which is necessary for the existence of a charged minimum (vacuum),
defines the interior of an ellipse in the $(\mu,\,\delta)$ plane:
\bear{c}
\fr{\mu^2}{a_1^2} + \fr{\delta^2}{ a_2^2} < 1\,,\quad
\mbox{where}\;\;
a_1 =  \fr{\lambda_4 + \lambda_5} {\Lambda_{3+}}\,,\quad
a_2 = \fr{\Lambda_{3-}}{\Lambda_{3+}} \,.
\eear{chvacreg}
At $m^2(T)>0$ and within this ellipse we also have $y_1>0$, $y_2>0$, therefore
the charged minimum exists and realizes the vacuum.
If $\mu,\,\delta$ point lies outside this ellipse, the minimum is neutral.

\bu {\bf The $\pmb{ (\mu,\,\delta)}$ plane}.
Fig.~\ref{charged_dmplane} represents the first sheet of the $(\mu,\,\delta)$ plane
for different sets of $\lambda_i$ from sector IV.
The states of charged vacuum are located within the grey ellipse described by eq.~\eqref{chvacreg}.
The left and right plots represent the case when the point ${\cal P}_0$ is outside or inside the ellipse,
respectively.
One can see that the situation is completely the same as in the sCPv case (Fig.~\ref{delmuplane}),
with only difference that the ellipse here represents the charged vacuum instead of sCPv.

The shaded area covers all ``zero points'' on the $(\mu,\,\delta)$ plane, which crossed in the past the ellipse of the charge-breaking vacua
(points 2 at rays II in the left plot only). Such trajectories describe the following phase evolution the during Universe cooling down\\[3mm]
 \cl{{\it EWs $\to$ CPc $\to$ charged $\to$ CPc}} \cl{(three phase transitions of the second order).}\\[2mm]

We obviously disregard the cases when the ``zero point'' lies inside the charged vacuum ellipse.

In the case of ``zero point'' 4 on ray II, if the ``zero point'' is located on ray I,
during cooling down, the Universe undergoes one standard EWSB phase transition\ \ \ {\it EWs $\to$ CPc} (see sect.~\ref{figonlyEWSB}).

\begin{figure*}[htb]
\includegraphics[width=0.3\textwidth]{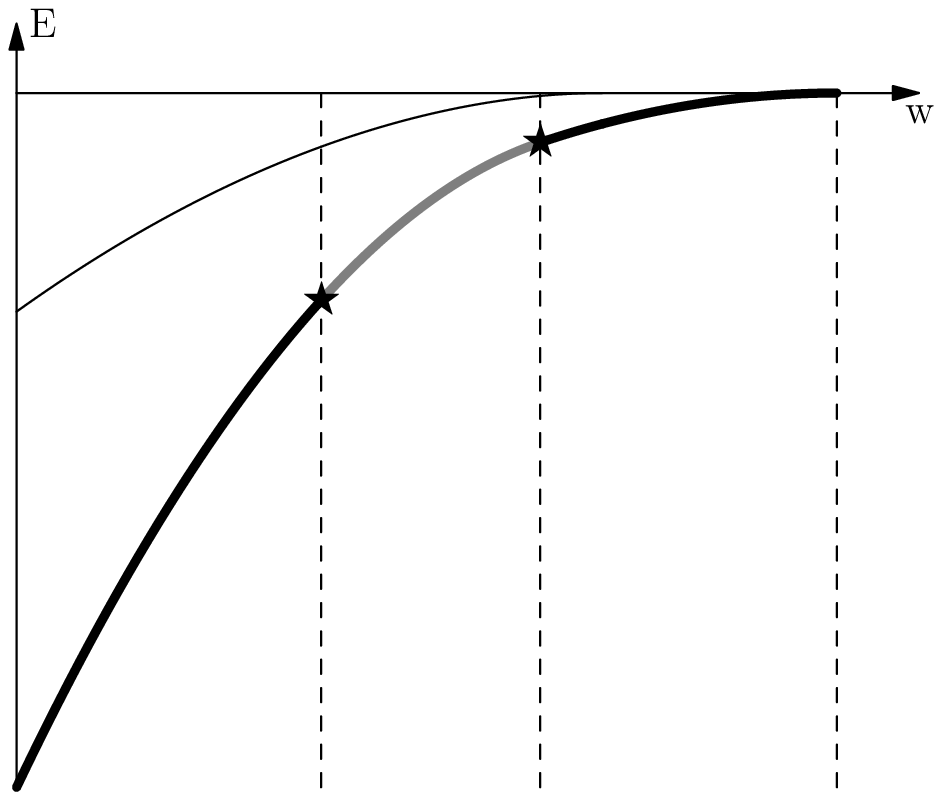}
\hspace{2mm}
\includegraphics[width=0.3\textwidth]{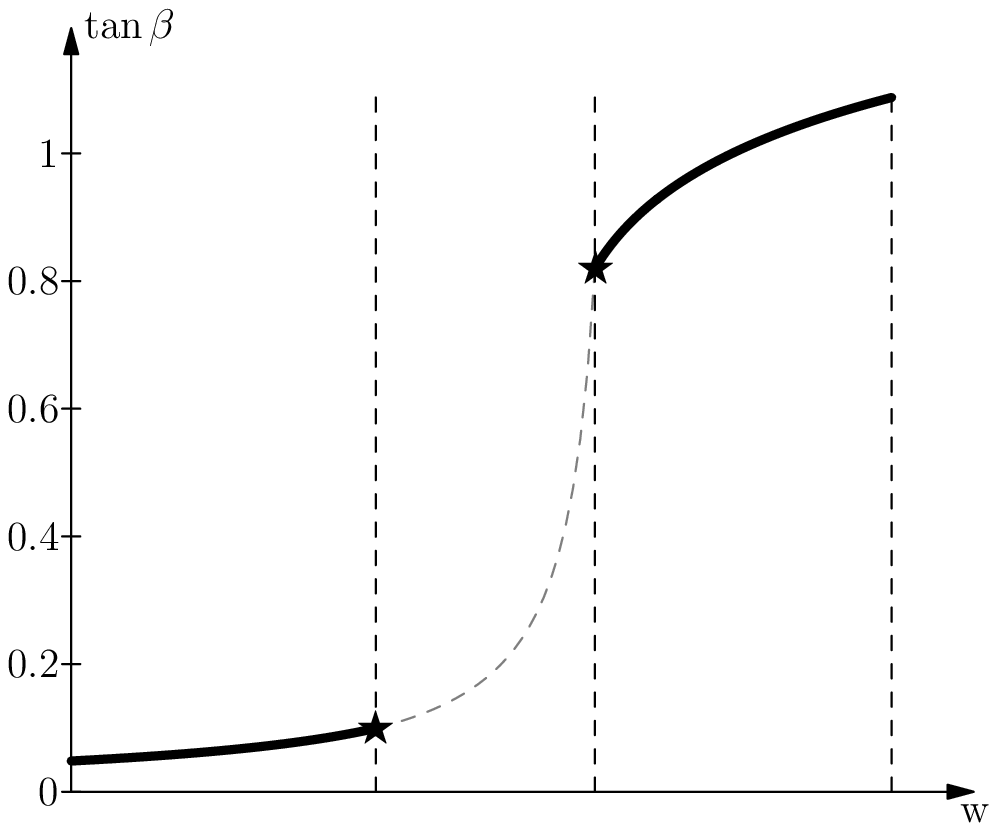}
\hspace{2mm}
\includegraphics[width=0.3\textwidth]{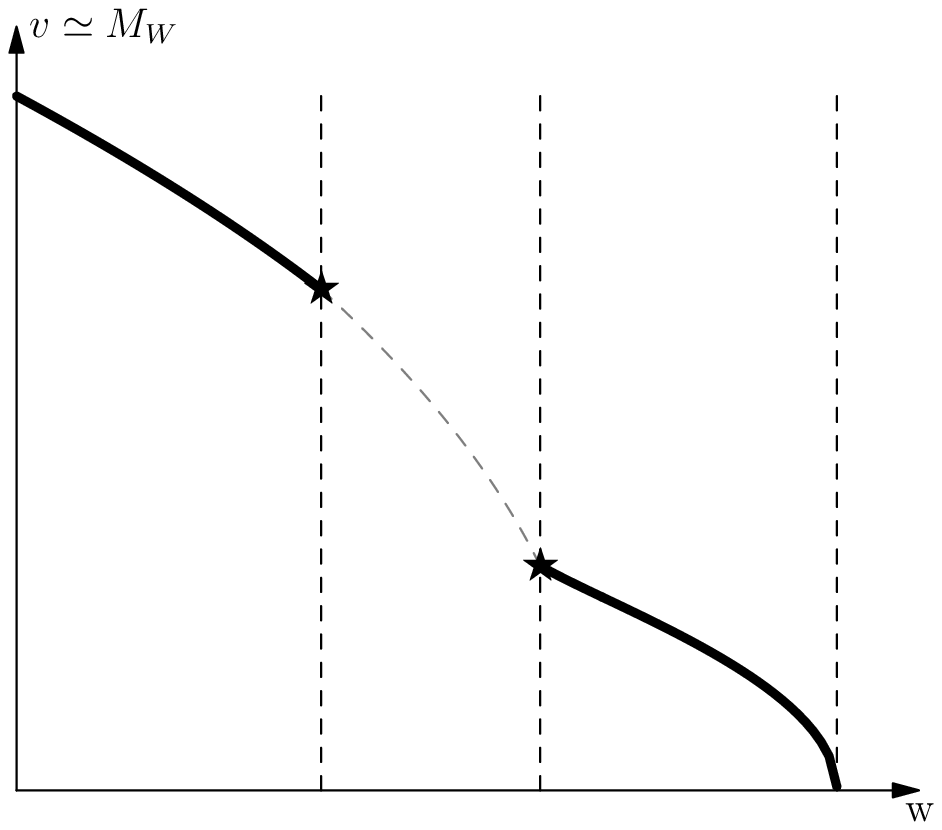}\\[5mm]
\includegraphics[width=0.3\textwidth]{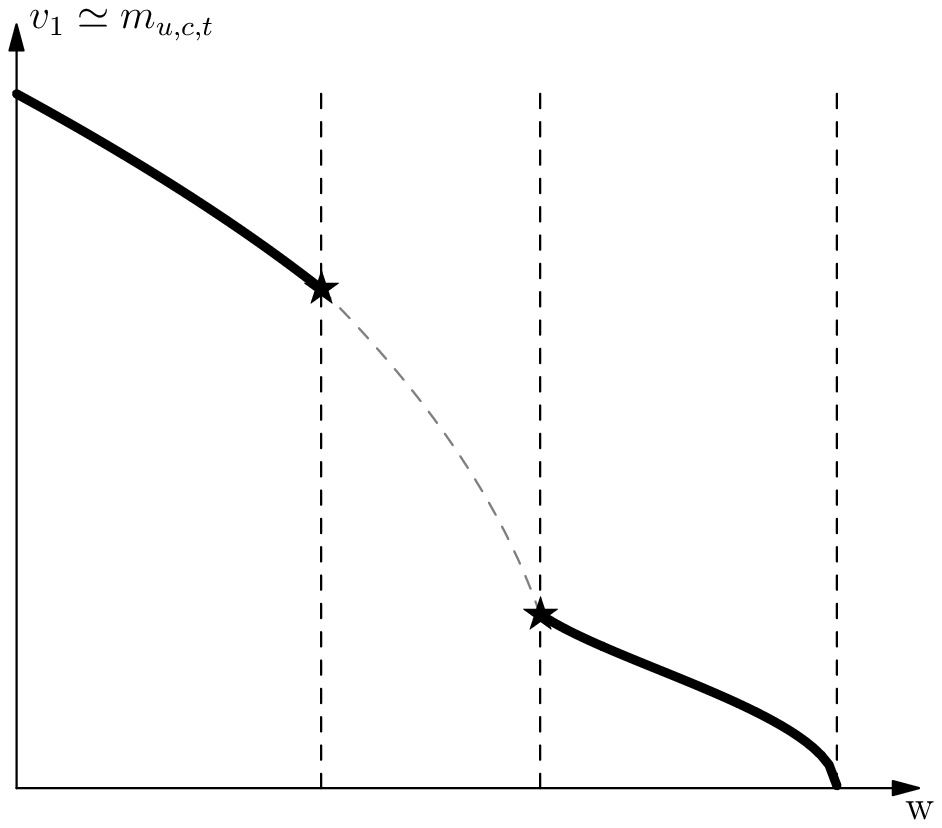}
\hspace{2mm}
\includegraphics[width=0.3\textwidth]{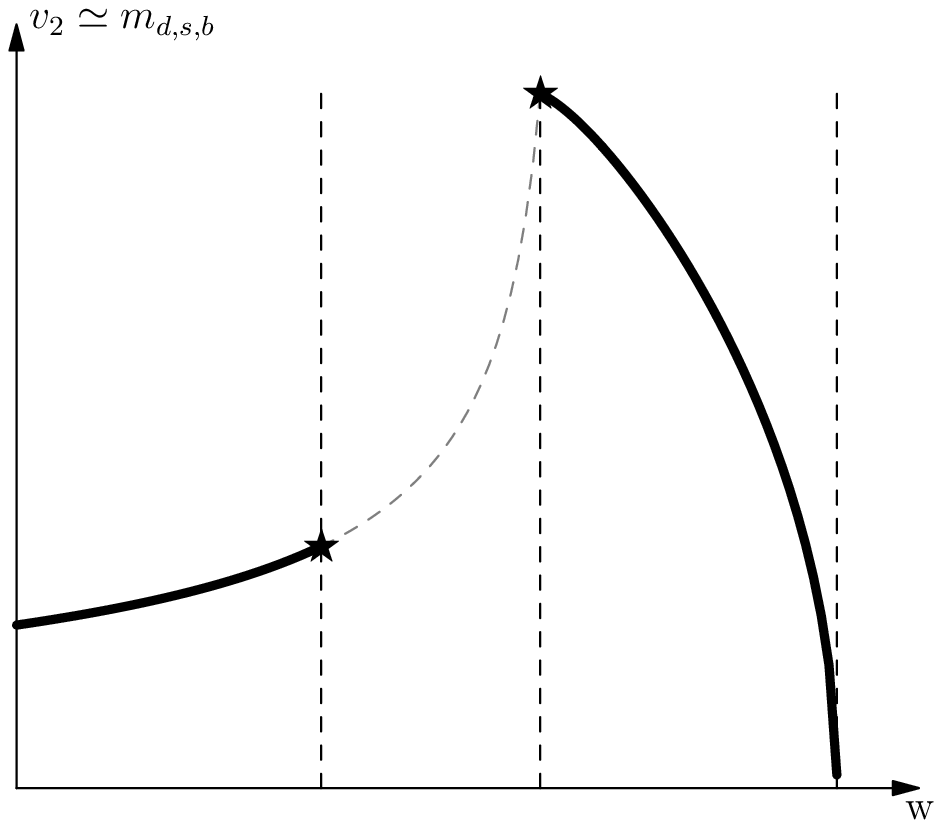}
\hspace{2mm}
\includegraphics[width=0.3\textwidth]{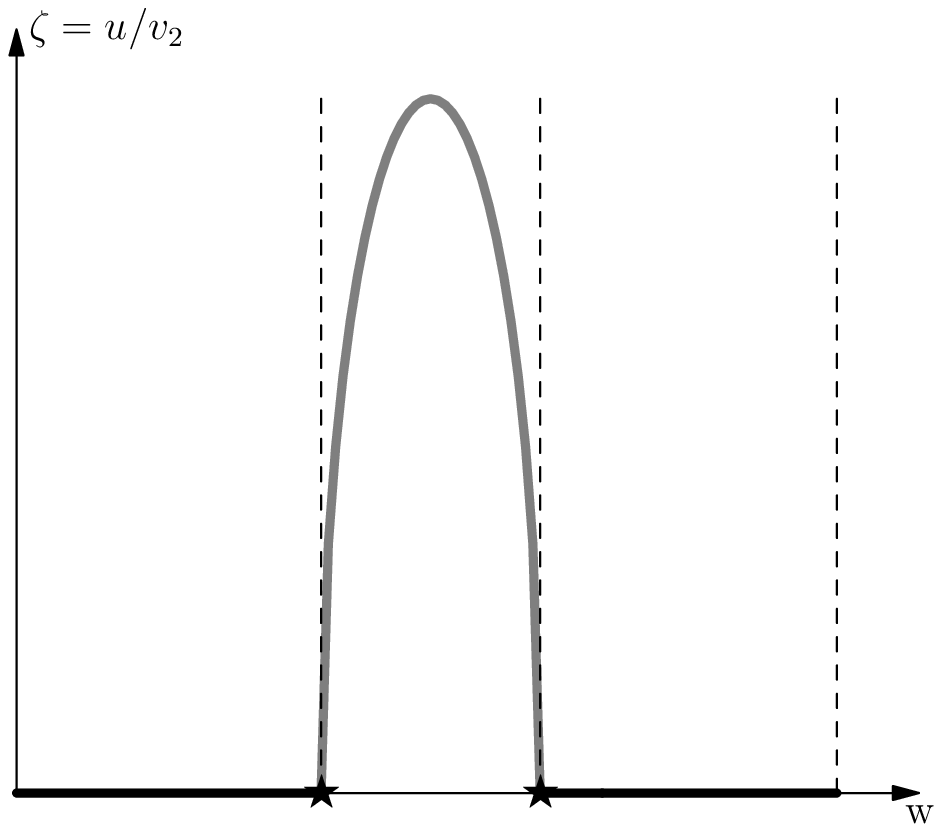}
\caption{\it Evolution of the $\zeta$ parameter and other physical quantities during transition through the charge-breaking vacuum.}
\label{chargedfigs}
\end{figure*}

\bu {\bf Evolution of the parameters of the vacuum state}.
A typical evolution of the vacuum parameters for ray I was already shown in Figs.~\ref{1CPcfigA}, Figs.~\ref{1CPcfigB}.
Fig.~\ref{chargedfigs} represents evolution of the parameters of the vacuum corresponding to ray II with ``zero point'' 2;
evolution from ``zero point'' 3 can also be read off these plots.
The notation and description of figures here are the same as for previous cases, e.g. in  Fig.~\ref{1stordermainA}.
The difference is that we do not show evolution of v.e.v.'s in the charged vacuum phase,
since these quantities have no clear physical sense in this phase.
Instead of that, in the right bottom plot we present temperature dependence for the order parameter in the charged vacuum phase,
which we choose as  $\zeta=\sqrt{Z/(y_3y_3^*)}=u/v_2$, i.e. the relative value of charge symmetry violating component
of v.e.v. $\la\phi_2\ra$. When the ray of physical states crosses the boundary of the ellipse \eqref{chvacreg},
this parameter starts growing from zero (Fig.~\ref{chargedfigs}), which is a typical behavior of an order parameter
of a second order phase transition.

The evolution of the v.e.v.'s near the charged vacuum phase demonstrates the same features as for the series of phase transitions
through the sCPv vacuum phase (Figs.~\ref{sCPvmainA},~\ref{sCPvmainB}).

\subsection{What if potential is not explicitly CP conserving?}

In Section~\ref{sectlagr} we presented arguments in favor of the softly $Z_2$ violating form of the Higgs potential.
In addition, to make discussion more specific, we limited ourselves to explicitly $CP$-conserving potentials \eqref{Z2potential}.
Here we briefly discuss what changes if the latter condition is lifted.

In this case a softly $Z_2$ violating potential has form \eqref{Z2potential} with complex $\lambda_5$ and $m_{12}^2$.
The rephasing transformation \eqref{rephas} allows one to
make $\lambda_5$ real and consider only $\mu$ complex.
This approach allows one to use the same subdivisions of $\lambda_i$ space into sectors as we used in this Section.

According to \cite{GK05}, \cite{GH05}, in this case the CPc phase does not exist.
All phase states violate the $CP$ symmetry, and we will denote them as CPv1, CPv2, etc.
Straight rays on the $(\mu,\,\delta)$ plane are replaced by straight rays in the 3-dimensional $(Re \mu,\,Im\mu,\,\delta)$ space.

The situation in {\bf sector IV} is analyzed in the same way as before.
With equations from \cite{GK07}, the condition for existence of the charged vacuum within the ellipse \eqref{chvacreg}
is transformed into a similar equation for 3-ellipsoid. Therefore, the phase evolution
is the same as discussed in sect.~\ref{seccharge}.

For other sectors a detailed study with explicit general equations for physical quantities and their behavior in different phases
becomes very complicated. In these cases the geometrical analysis of \cite{Ivan1} allows one to obtain general picture of phase transformations.
In {\bf sector II} and in the case of no other phase transitions except EWSB, we obtain roughly the same picture as above,
sect.~\ref{figonlyEWSB},~\ref{sec1ordtr} with a natural change
of the labels: CPc$\to$CPv, CPv1, CPv2. In {\bf sector III} the picture changes,
as the doubly degenerated sCPv states disappear. The ray of physical states goes now in 3D and is inclined towards the plane,
where an ellipse of doubly degenerate states similar to sCPv exists.
The ray can traverse this plane, resulting in a first order phase transition, like the one in our sector III,
with a nonzero specific heat that depends on the slope of this ray.
Nevertheless, the region of all modern values of $(\mu,\,\delta)$, which would go through the ellipse at high temperature,
can be described in an essentially similar way as the shaded area in Fig.~\ref{delmuplane}.

\section{Discussion}

We considered here the phases and thermal phase transitions in the 2HDM.
We limited ourselves with the case of softly broken $Z_2$ symmetry,
presenting arguments in favor of its realization in Nature.
Besides, in what concerns sequences of phase transitions this case is representative of the most general situation \cite{Ivan1}, obtained in tree approximation.
We obtained a rich picture of possible phase states and phase transitions. Taking into account finite mass and higher order effects
will modify the simple temperature dependence of effective potential parameters discussed in sect.~\ref{secdep}.
We believe that with these modifications the picture of possible phase states and phase transitions can only become richer.In particular,  some  second order transition can be transformed into the first order one.
However even the picture obtained here looks very interesting.

Let us now discuss some general features of the picture obtained focusing on the case of the explicitly $CP$ conserving potentials.

{\bf Possible sequences of phase states}.

Thermal evolution can be split into two phases whose properties are rather decoupled from each other.

At the extremely high temperature the vacuum state can be either EW symmetric or EW-violating, but $CP$-conserving (CPc).
In the former case, which we see as the most probable, the Universe during its cooling down passes
to a CPc phase (EWSB phase transition at $T=T_{EW}$, Fig.~\ref{figewsb}, left plot).
In the second case, cooling down either keeps Universe in the CPc phase for a long time or brings it into and then out of
the EW symmetric phase, that is, first restoring and then breaking again the EW symmetry.

As the temperature goes down, the system enters the lower-temperature stage of its evolution,
which in our analysis corresponds to a transition from the second to the first sheet of the $(\mu,\,\delta)$ plane.
The mass parameters of the system keep changing,
and the phase state of system can either vary continuously or exhibit one or two additional phase transitions.
In these transitions our system can either evolve through a charged vacuum phase (with very unusual properties),
or cross the sCPv phase, or stop in the sCPv phase state (in this case one should observes spontaneous $CP$-violation in the Higgs sector in today's world)
or finally cross the line of the $k$-symmetry of the CPc phases.
In our approximation only the transition CPc$\to$CPc through the $k$-symmetric state (at $k\neq 1$) is of the first order.
We therefore list all possible sequences of phase states indicating in all cases the type of phase transition:

\begin{widetext}
\be
\left.\begin{array}{ll}
(a) & EW\\
\\
(b) & CPc\\
\\
(c) & CPc\;\xrightarrow{II}\;EW\;\;\end{array}\right\}\;\xrightarrow{II}\;
\begin{array}{ll} CPc & (I)\\
 CPc \xrightarrow{II}\; charged \; \xrightarrow{II} CPc& (II)\\
 CPc \xrightarrow{I} CPc2 & (III)\\
  CPc \xrightarrow{II} sCPv &(IV)\\
CPc \xrightarrow{II} sCPv \xrightarrow{II} CPc & (V)
\end{array}\label{sequence}\ee
\end{widetext}

We obviously omitted from this list the sequence (EW $\xrightarrow{II}$ CPc $\xrightarrow{II}$ charged), which disagrees with today's Universe.
Each possible sequence from the left column can be combined with any possible sequence from the right column.
In the case (b)+(I), the history of the Universe contains no phase transition at all.\\

{\bf Regions of parameters allowing for different paths of  phase evolution.}

We presented a method how to describe regions in the space of the zero-temperature parameters of the model that lead
to each specific type of thermal evolution of the Universe
(see the shaded areas in Figs.~\ref{fig1delta},~\ref{delmuplane}).
To cast them into the corresponding regions of observables, such as masses and couplings constants,
is a natural task for continuation of this work.\\

{\bf Rearrangement of particle mass spectrum}.
In most examples considered here the value of $\tan\beta$ changes strongly. It jumps at the first order phase transition
and shows a continuous but very sizable evolution for the other cases.
In most of our examples we saw $\tan\beta$ increase towards high temperatures, i.e. in the past,
but a simple change of the basic Higgs fields $1\leftrightarrow 2$ leads to $\tan\beta\to 1/\tan\beta$, that is, to decreasing $\tan\beta$.
In the latter case equations \eqref{massrelmot} shows that the fermion mass spectrum within one generation
can be rearranged. Under these circumstances, it is possible that in the past the decay $t\to Wb$ was suppressed,
and $W\to tb$ decay was allowed or even that the $b$-quark was heavier than $t$-quark.

In a sequence of several phase transitions, it is only the first one, the EWSB, that must take place at the electroweak temperature scale.
It might happen that the other phase transitions take place at much lower temperatures. It means that there exists a possibility that the last phase transition took place relatively lately in the history of the Universe. If this is the case, then the possible rearrangement of the quark mass spectrum could have even more spectacular effects.
For example, if the Universe lived long enough in an intermediate phase with $m_d<m_u$, then the proton could be lighter than the neutron and could even decay into it during this intermediate stage, which has profound cosmological consequences. More delicate, even weak variation of  relation between hadron masses at low temperatures (for example, within some stars) can influence for process of nucleosynthesis.

Another interesting opportunity, which can be realized in many cases, is a non-monotonic dependence of masses of particle on temperature,
when they start from zero at EWSB phase transition,
grow and overshoot their today's values and drop down after subsequent phase transitions.

In addition, the rearrangement of fermion masses can be viewed as yet another phase transition in fermion subsystem, with own fluctuations, etc. The study of this possible phase transition goes beyond the approach developed in this paper.\\

{\bf Possible relations to cosmology}.

1. Different phenomena discussed here can give rise to new effects
in the structure of Cosmic Microwave Background radiation and other cosmological observables. Feasibility of their observation is a subject for future studies.

The cases with new phase transitions in addition to the standard EWSB lead to additional stages in the early history of the Universe
with strong fluctuations near the phase transition points. For example, in many cases, see Figures~\ref{delmuplane},~\ref{fig1delta},
there exists either a meta-stable local minimum state or other extrema just above the vacuum state. Possible virtual transitions to these states can enhance fluctuations and their observable effects.

2. If the charge-breaking vacuum state was indeed an intermediate stage of the evolution of the Universe, then a number of unexpected effects appear and they can strongly influence the modern situation. First of all, in the charge-breaking phase all the gauge bosons are massive, and electric charge is not conserved. Also, in this phase the long ranged forces, like electromagnetism in our world, are absent, and the only long range force is gravity. So, the local electric neutrality of medium can be strongly violated. After a phase transition to the modern charge-conserving vacuum, one could have strong deviations from the average electroneutrality, originating from the charge-breaking vacuum phase. These are not the standard charge fluctuations, and they will result in a strong relative motion of separate parts of the Universe, which can result either in strong mixing and averaging of matter or in production of structures like caustics (proto-galaxies). The restoration of the electric neutrality can go on during a very long time after the phase transition to neutral vacuum.

3. In the standard approach  the temperature of the phase transition is unavoidably set by the electroweak scale.
In our model the same is valid for the first EWSB transition.
However, the temperature of the last phase transition can be sufficiently low, that is, the ``zero point'' in Figs.~\ref{fig1delta},~\ref{delmuplane}, can be close to the phase separation line. Certainly, for a detailed description
of such situation our approximation must be improved.\\

{\bf Acknowledgments.} We are thankful to M.~Krawczyk, M.~Maniatis, S.~Kanemura, L.~Okun,  V.~Rubakov, R.~Santos, A.~Slavnov for useful discussions. This research was supported by Russian grants RFBR 08-02-00334-a and NSh-1027.2008.2. The work of I.P.I. was supported also by the Belgian Fund F.R.S.-FNRS via the contract of Charg\'e de recherches.\\

\begin{widetext}
\appendix
\section{Temperature corrections to potential for the case of hard violation of $Z_2$ symmetry}\label{Temphard}

In the general case of hard violation of $Z_2$ symmetry  all mass terms have non-zero corrections.
In addition to the usual one loop diagrams of form of Fig.~\ref{figtadpolediag},
one adds diagrams with a specific bilinear vertex generated by the $\vak$ term. We present the resulting mass corrections
for the case of explicitly $CP$-conserving potential assuming field renormalization
$\phi_i\to \phi_i/\sqrt{1-\vak^2}$, taking into account a number of $\vak$ vertex insertions in each line.
Instead of eq.~\eqref{Tempdep} we obtain after simple algebra
\bear{c}
2c_1=3\lambda_1+2\lambda_3+\lambda_4+6\vak(2\lambda_6+\lambda_7)+
\vak^2(6\lambda_3+3\lambda_4+3\lambda_2)+6\vak^3\lambda_7\,,\\
2c_2=3\lambda_2+2\lambda_3+\lambda_4+6\vak(\lambda_6+2\lambda_7)+
\vak^2(6\lambda_3+3\lambda_4+3\lambda_2)+6\vak^3\lambda_6\,,\\
2c_{12}=[6(\lambda_6+\lambda_7)+\vak(\lambda_3+2\lambda_4+\lambda_5)](1+\vak^2)
+\vak(3\lambda_1+2\lambda_2+4\lambda_3+2\lambda_4)\,.
  \eear{Tempdeph}
Here we include the diagram factor in definition of $\vak$.

These equations show similar to that it was shown in sect.~\ref{secdep}
that {\it the curve of physical states} is now a straight ray in the in the 4-dimensional space $(m_{11}^2\,,
m_{22}^2\,,Re m_{12}^2\,,Im m_{12}^2)$ and consequently in the 3-dimensional space $(Re\mu\,,Im\mu,\,\delta)$ (definition of quantities $Re \mu$,  $Im \mu$ is evident in analogy with \eqref{potparam}).

\section{Decomposition of the potential around arbitrary EWv extremum}

We present here  useful equation \cite{GK07} for decomposition of the potential around arbitrary EWv extremum $N$, obtained with the aid of eq.~\eqref{vacEy}. It is valid  for the most general potential \eqref{potential}. The complexity of the potential in this form is concentrated in the values $\lambda_{5,6,7}$ and $y_3$.
 \bear{c}
V={\cal E}_N^{ext}+V_4(x_i-y_{i,N}) +{\cal R}\cdot{\cal D}(\phi,N)\,;\\[2mm]
{\cal D}(\phi,\,N)\!=\!x_1y_2+x_2y_1-x_3y_{3^\dagger}-y_3x_{3^\dagger}\equiv\\[2mm]
\equiv (\phi_1\la\phi_2\ra_N\!-\!\phi_2\la\phi_1\ra_N)^\dagger
(\phi_1\la\phi_2\ra_N\!-\!\phi_2\la\phi_1\ra_N)\!\,;\\[2mm]
 \begin{array}{c l}
 {\cal R}=0&\;\;\mbox{\it for charged  extremum}\,\\[1mm]
\left.  {\cal R}=\fr{M_{H^\pm}^2}{y_1+y_2}\right|_N&\;\;\mbox{\it
for neutral extremum}\;N\,.
  \end{array}\end{array}
\label{decomp}\ee
Here $M^2_{H^\pm}$ is the squared mass of charged Higgs boson (it can be negative if considered extremum is not minimum of potential).
The quantity $\cal{D}(\phi,\,N)$ can be treated as the
the ``distance'' between some set of fields and extremum $N$ with $\lr{\phi_i}=\lr\phi_{iN}$.
By construction $\lr{{\cal D}}\ge 0$ for any classical values of $\phi_i$.

\section{Parameters for Figures}
All the figures, showing evolution of physical parameters are obtained by numerical calculations.
Here we present the parameres, used for the calculations:
\begin{center}
\begin{tabular}[c]{|l||l|l|l|l|l||l|l|}
	\hline
	Figure 		    &$\lambda_1$&$\lambda_2$&$\lambda_3$&$\lambda_4$&$\lambda_5$&$\mu$ &$\delta$ \\ \hline
	Fig.~\ref{1CPcfigA} &   0.076   &   1.22    &   1.69    &   0.275   &   -1.235  & 0.39 &  -0.17  \\ \hline
	Fig.~\ref{1CPcfigB} &   0.222   &   3.556   &   1.11    &   0.205   &    0.125  & 0.12 &   0.02  \\ \hline
Fig.~\ref{1stordermainA}    &   0.488   &   2.475   &   0.9     &  -0.635   &   -0.125  & 0.175&  -0.17  \\ \hline
Fig.~\ref{1stordermainB}    &   0.222   &   3.556   &   1.12    &   0.208   &    0.428  & 0.12 &   0.014 \\ \hline
	Fig.~\ref{sCPvmainA}&   0.197   &   3.16    &   1.21    &   0.6     &    1.08   & 0.065&  -0.05  \\ \hline
	Fig.~\ref{sCPvmainB}&   0.22    &   3.556   &   1.12    &   0.218   &    0.418  & 0.15 &   0.014 \\ \hline
Fig.~\ref{chargedfigs}      &   0.217   &   3.48    &   1.13    &   6.68    &   -5.32   & 0.145&  -0.18  \\ \hline
\end{tabular}
\end{center}

\end{widetext}

\end{document}